\documentclass[structabstract]{aa}
\usepackage{graphicx}
\usepackage{txfonts}
\def\ltsim{\hbox{\raise 2pt \hbox {$<$} \kern-1.1em \lower 4pt \hbox {$\sim$}}}
\def\ltapprox{\hbox{\raise 2pt \hbox {$<$} \kern-1.1em \lower 5pt \hbox
{$\approx$}}}
\def\gtsim{\hbox{\raise 2pt \hbox {$>$} \kern-1.1em \lower 4pt \hbox {$\sim$}}}
\def\gtapprox{\hbox{\raise 2pt \hbox {$>$} \kern-1.1em \lower 5pt \hbox
{$\approx$}}}
\def\skuno{\vskip 20pt}

\begin{document}

   \title{Radio halos in nearby (z $<$ 0.4) clusters of galaxies}

   \author{G. Giovannini \inst{1,2}, A. Bonafede \inst{1,2}, 
           L. Feretti \inst{2}, 
           F. Govoni \inst{3}, M. Murgia \inst{2,3}, F. Ferrari \inst{2},
           G. Monti \inst{2}}

 \institute{(1) Dipartimento di Astronomia, via Ranzani 1, 40127 Bologna, I \\
            (2) Istituto di Radioastronomia-INAF, via P.Gobetti 101,
                40129 Bologna, I \\
            (3) Osservatorio Astronomico di Cagliari - INAF, 
                Strada 54, Loc. Poggio dei Pini, 09012 Capoterra (Ca), I
             }

\authorrunning{Giovannini et al.}

\abstract{The Intra-Cluster Medium is characterized by thermal emission, 
and by the presence of large scale magnetic fields. In some clusters of 
galaxies a diffuse non-thermal emission is also present, located at the 
cluster center and named radio halo. These sources indicate the existence
of relativistic particles and magnetic fields in the cluster volume. }
{In this paper we collect data on all known nearby cluster radio 
halos (z $<$ 0.4),
to discuss their statistical properties and to investigate their origin.}
{We searched for published data on radio halos and reduced new and archive VLA
data to increase the number of known radio halos.}
{We present data on 31 radio halos, 1 new relic source, and 1 giant 
filament. We note the discovery of a small size diffuse radio emission
in a cluster (A1213) with very low X-ray luminosity.
Among statistical results we confirm the correlation between the 
average
halo radio spectral index and the cluster temperature. We also discuss the
high percentage of clusters where both a relic and a radio halo is present. 
}
{The sample of radio halos discussed here represents the population of radio 
halos observable with present radio
telescopes. The new telescope generation is necessary for a more detailed
multifrequency study, and to investigate the possible existence of a 
population of radio halos with different properties.}

\keywords{galaxies:cluster:non-thermal}

\maketitle
\section{Introduction}

The baryonic content of galaxy clusters
is dominated by the hot (T $\sim$ 2 −- 10 keV) intergalactic gas
whose thermal emission is observed in X-rays. Thermal emission is a common
property of all clusters of galaxies and has been detected even in
poor galaxy groups as well as in optical filaments connecting
rich clusters.

The Intra-Cluster Medium (ICM) is also characterized by the presence of
large scale magnetic fields.  A magnetized plasma between an observer and 
a radio source changes the observed properties of the polarized emission 
from the
radio source. Therefore information on clusters 
magnetic fields can be determined, in conjunction
with X-ray observations of the hot gas, through the analysis of the
Rotation Measure (RM) of radio galaxies in the background or in the
galaxy clusters themselves. Detailed RM studies have shown
that in the ICM large scale magnetic fields at microgauss level (higher
in central cooling cluster regions) are present (e.g. \cite{govf04}, 
\cite{bon09b}).

In some clusters of galaxies a diffuse non-thermal emission is also present
(e.g. \cite{fer05a}, \cite{ferra08}).
These sources indicate the existence of relativistic particles and
magnetic fields in the cluster volume, thus the presence of non-thermal
processes in the hot ICM.  

These large synchrotron sources,
which are not obviously associated with any individual galaxy,
have been identified as relics, mini-halos, and
halos. They are diffuse, low-surface brightness 
($\simeq 10^{-6}$ Jy/arcsec$^2$ at 1.4 GHz), 
steep-spectrum\footnote{S($\nu$)$\propto \nu^{- \alpha}$} ($\alpha > 1$) 
sources, and represent the best evidence for the presence 
of large-scale magnetic
fields and relativistic particles in the cluster peripheral regions (relics,
\cite{gf04}, \cite{bon09a}),
in the central regions of relaxed clusters (mini-halos, \cite{gov09}),
and in non relaxed
clusters (halos, \cite{fer04a}). In this paper we will focus on radio halos. A
similar review on radio relics is in progress, and we refer to recent papers
as \cite{gov09} and \cite{mur09} about mini-halos. 

Radio halos are faint, steep-spectrum sources located at the cluster
center (e.g. \cite{fg08}). Their radio emission is typically unpolarized
with the exception of A2255 (\cite{gov05}), and MACSJ0717.7+3745 
(\cite{bon09b}).
Radio halos are typically found in clusters which show
significant evidence of an ongoing merger (e.g. \cite{buo01}, \cite{gov04}). 
It has been proposed that recent cluster
mergers may play an important role in the re-acceleration of the
radio-emitting relativistic particles, thus providing the energy
to these extended sources (e.g. \cite{bru01}, \cite{pet01}).

\begin{figure*}
\centering
\includegraphics[width=18cm]{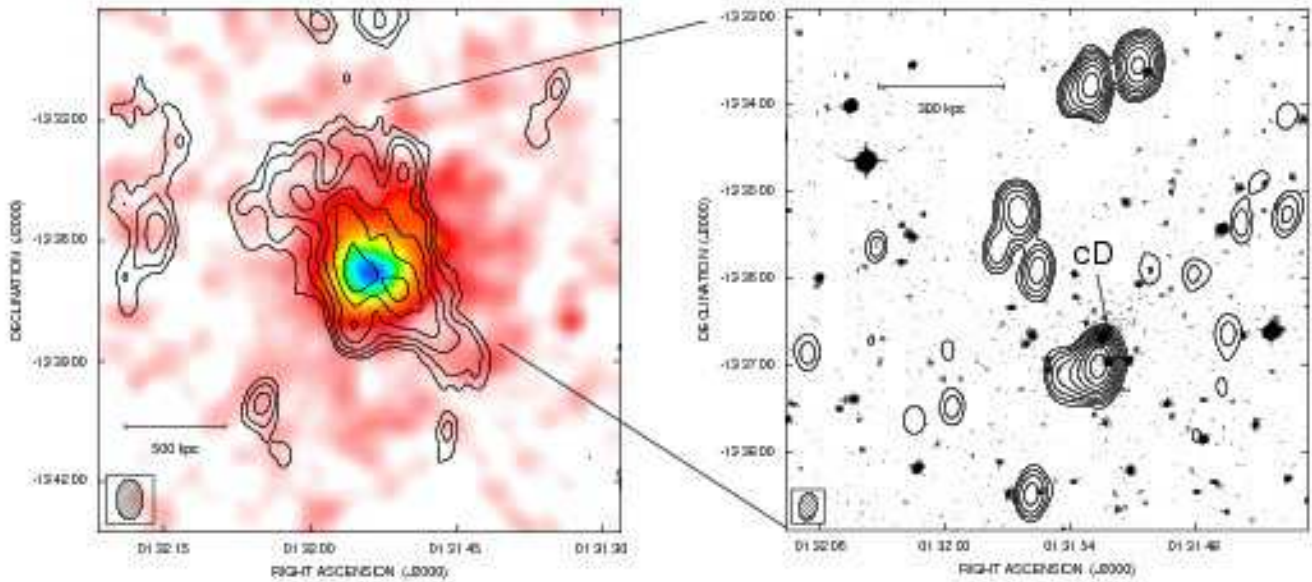}
\caption{
Left: radio contours of the extended halo in A209 obtained with the VLA at 
1.4 GHz combining data in C and D configuration, after subtraction of 
discrete sources.
The HPBW is $60'' \times 40''$ (PA 0$^\circ$), 
and the noise level is 0.05 mJy/beam.
The first contour level is drawn at 0.2 mJy/beam and the others
are spaced by a factor $\sqrt2$. 
We note on the East of the radio halo a faint diffuse emission in N-S 
direction. The shape and extension suggest its identification with a possible
relic source, however because of the low brightness and of the presence
of optical candidates further observations are necessary to confirm this.
The contours of the radio intensity
are overlaid onto the Rosat HRI X-ray image in the 0.1-2.4 keV band. 
The X-ray image has been smoothed with a Gaussian of $\sigma=16''$.
Right: radio contours obtained with the VLA at 1.4 GHz in C configuration 
of A209.
The HPBW of the radio image is $20.1'' \times 12.8''$ (PA -3$^\circ$),
and the noise level is 
0.03 mJy/beam. The first contour level is drawn at 0.07 mJy/beam and the others
are spaced by a factor 2. The contours of the radio intensity are 
overlaid onto the optical red image obtained from the SERC/ESO surveys. 
The central cD galaxy 
(pointed out by a line) is marginally 
visible on the top of the head-tail radio galaxy. Only discrete sources
are visible because of the missing of short baselines. Here and in the 
following figures the restoring HPBW is shown in the bottom left corner.}
\label{a209} 
\end{figure*}

About 20 radio halos are known from literature data up to now 
(e.g. \cite{fg98}, \cite{fer00}, \cite{gov01a}, \cite{bac03},
\cite{ven07}, \cite{ven08}, and \cite{ven09}), 
and their properties have been discussed in different papers.
The knowledge of the properties of these sources has increased
significantly in recent years, due to higher sensitivity radio images and to
the development of theoretical models.  The importance of these
sources is that they are large scale features, which are related to
other cluster properties in the optical and X-ray domain, and are thus
directly connected to the cluster history and evolution.

Here we will present new data and images of radio halos and
collect all data published on halos for a statistical study.
The organization of this paper is as follows:
in Sect. 2 we present new data on nearby (z$<$ 0.4) radio halos (a study 
of the extended emission in rich clusters at redshift $>$ 0.4 is in progress,
see e.g. \cite{bon09b}); in Sect. 3 we present new radio halos
and report in Table 2 relevant
data on all radio halos in nearby clusters known up to date. 
Correlations and statistical
results are discussed in Sect. 4 and Conclusions are reported in Sect. 5.

The intrinsic parameters quoted in this paper are computed for 
a $\Lambda$CDM cosmology with $H_0$ = 71 km s$^{-1}$Mpc$^{-1}$,
$\Omega_m$ = 0.27, and $\Omega_{\Lambda}$ = 0.73.

\section{Observations and data reduction}

Here we present new data for radio halos obtained with the VLA at 1.4 GHz.
The list of the clusters, together with the observation position, 
the on source time,
the VLA configuration, and the observing date is reported in Tab. 1.
Information on final images (angular resolution and noise level) are
reported in the figure captions. All new data are from pointed observations, 
and include proprietary data as well as VLA public archive data.\\

\begin{table*}
\caption{Observing parameters}
\label{tab1}
\centering
\begin{tabular}{lllccc}
\hline\hline
Cluster&  RA       & DEC   & Obs. Time & Array & Date    \\
       &  J2000    & J2000 & minutes   &       & dd-mm-yy  \\
\hline
A209   & 01 31 57.00 & $-$13 34 35.0 &   20 & C &  15-APR-2004  \\
       &             &               &   90 & D &  03-JUL-2004  \\
A521   & 04 54 12.00 & $-$10 15 00.0 &   360& B/C& 22-SEP-2002  \\
A697   & 08 42 57.76 & $+$36 21 45.3 &   50 & C &  30-MAR-1996   \\
A851   & 09 42 48.60 & $+$47 00 00.0 &   20 & C &  15-MAR-1995 \\
       &             &               &   70 & D &  03-APR-2000 \\
A1213  & 11 16 40.00 & $+$29 16 00.0 &   90 & C &  06-MAR-2008 \\
       &             &               &   90 & C/D& 02-JUN-2008 \\
A1351  & 11 42 32.50 & $+$58 32 08.0 &   15 & D &  15-MAR-1995 \\
       &             &               &  120 & C &  03-APR-2000 \\
A1758  & 13 32 32.00 & $+$50 30 36.0 &  150 & D &  11-MAR-2003 \\
       &             &               &  150 & C &  06-MAY-2004 \\
A1995  & 14 52 47.50 & $+$58 03 08.0 &  120 & C &  03-APR-2000 \\
       &             &               &  15  & D &  15-MAR-1995  \\
A2034  & 15 10 17.00 & $+$33 31 00.0 &  30 & C &  06-AUG-2001  \\
       &             &               &  30 & D &  24-MAR-2003 \\
A2294  & 17 23 13.80 & $+$85 53 21.0 &  15 & D &  15-MAR-1995 \\
       &             &               &  60 & C &  01-APR-2000 \\
A3444  & 10 23 50.11 & $-$27 15 15.8 &  240 & A/B& 02-OCT-2003 \\
       &             &               &  240 & C/D& 29-MAY-2004 \\
CL1446+26&14 49 28.70& $+$26 07 54.10&  120 & C &  01-APR-2000   \\
\noalign{\smallskip}
\hline
\multicolumn{6}{l}{\scriptsize Col. 1: Cluster Name; Col. 2, Col. 3: 
Observation pointing  (RA J2000, DEC J2000);}\\
\multicolumn{6}{l}{\scriptsize Col. 4: On source observing time; 
Col. 5: VLA configuration; Col. 6: Observing dates.}\\
\end{tabular}
\end{table*}

To derive final images we calibrated the data with the standard technique 
using the AIPS package. Calibrated data were carefully edited and 
self-calibrated 
in phase and gain. To obtain images and flux densities of the
extended halo sources, unrelated discrete sources have been subtracted in
the uv-plane. 

To this aim we produced high resolution images with a uniform weight 
(ROBUST = -5), and without the short baselines. In these images all discrete
sources are present, but the extended diffuse sources (halos or relics) are
absent because of the missing of short spacings in the u-v plane.

We checked that the total flux density in the clean components is in agreement
with the total flux density visible in the u-v plane. We then 
selected the clean
components of sources to be subracted (usually in the central region)
using the task CCEDT and the AIPS procedure BOX2CC. After a comparison among 
source flux densities in clean components and from a gaussian fit (task JMFIT),
clean components were subtracted from the u-v data. With the new data set we 
produced images with uniform and natural weights and
carefully compared the total flux density present in the short baselines
with the total flux in the clean components. Finally we derived the 
parameters of the diffuse sources.

The error of the flux density of diffuse radio sources 
has been obtained as due to the combination of the noise level (derived from
the individual images), and the uncertainty due to the flux density calibration
(estimated of the order of 3$\%$). In addition, in dealing with extended
sources, one has to take into account the presence of possible variations
in the zero-level of the image. We estimated the zero-level in the images
in region free of sources around the radio halos using the AIPS task 'IMEAN',
and found that it is very good, and does not affect the measured halo 
flux density.
Finally, the accuracy of the flux density estimate is related to the 
subtraction of discrete sources.
We estimated this last point comparing the flux density in clean 
components subtracted in u-v data with the flux density estimated with
a gaussian fit.
Final flux density uncertainties are reported in Table 2.
For published halos we report the published flux density uncertainty when 
available. If not published we estimated it as discussed before.

\section{New radio halos at 1.4 GHz}

We present here new images of radio halos in 10 different clusters 
(see Tab. 1). Moreover we present new data on two radio halo candidates, 
which in the light of new data we classify as a giant filament (Abell 3444)
and a new relic source (CL1446+26).
We found that in A697 (\cite{ven08}) and A521 (\cite{bru08})
the diffuse emission previously found only at lower frequency is also clearly 
visible at 1.4 GHz (see also \cite{dal09}). 
Moreover we present improved images at 1.4 GHz for 
A209, and A1758 previously discussed at this frequency only in conference
proceedings. For the other clusters we confirm with pointed observations
the presence of a diffuse radio emission.
Details on the individual sources are given below.
\skuno
\noindent
{\bf Abell 209} is a rich cluster of galaxies at z = 0.206.
It is dominated by a central cD galaxy, and the peak of 
the cluster mass distribution is close to this galaxy. 
A candidate 
gravitational arc is visible inside the cD halo (\cite{dah02}).
ROSAT HRI data
for this cluster indicate an irregular X-ray morphology with
significant substructure (\cite{riz98}).
X-ray data and optical density distribution suggest the presence of a cluster
merger (\cite{mer03}).
The projected mass distribution estimated with the weak lensing technique
(\cite{p-h07}), 
shows a pronounced
asymmetry, with an elongated structure extending from the SE to the NW. 
A similar elongation was previously detected in the X-ray
emission map by \cite{riz98}, and it is confirmed by the distribution of 
galaxy colours (\cite{p-h07}).

In the radio band an extended halo is detected. An image at 1.4 GHz was first
presented in \cite{gio06}.
\cite{ven07} presented a radio image at 610 MHz obtained with the
GMRT but the smaller angular size and low flux density with 
respect to the 1.4 GHz data, suggest some missing flux in GMRT data.

In Fig. \ref{a209} we show the central cluster radio emission
compared with optical and X-ray images.
On the left we present the radio contours of the extended halo source 
obtained by
combining C and D configuration data.
Discrete radio sources have been subtracted as discussed before. 
The halo total flux density is 16.9 mJy
and the largest size is $\sim$ 7$'$ in PA $\sim$ 45$^\circ$.
On the East of the radio halo an extended elongated structure is present at 
about 4 $\sigma$ level, possibly identified as a peripheral relic 
radio source. A deeper
VLA image is necessary to confirm the detection and to study it
in more detail.
For a comparison between the radio and the X-ray cluster emission,  
the contours of the radio intensity are overlaid 
onto the Rosat HRI X-ray image in the 0.1-2.4 keV band.
On the right we present a zoom of a full resolution image obtained with 
the VLA in 
the C-array 
of the cluster center overlaid onto an optical image.
The optical images herein were taken  
from the 
Optical Digitized Sky Survey \footnote{See http://archive.eso.org/dss/dss}.
Because of the angular resolution and the 
missing of short baselines only discrete sources are visible. The central
cD galaxy appears to be radio quiet, but a head-tail radio galaxy is 
visible near the cluster center.\\

\noindent
{\bf Abell 521} at z = 0.2533 has been studied in detail in the optical and  
X-ray using Chandra observations (\cite{ferra06}). A521 is a spectacular
example of a multiple merger cluster made up of several substructures 
converging at different epochs towards the centre of the system. The very 
perturbed dynamical state of this cluster is also confirmed by the discovery 
of a radio relic in its South-East region, using VLA observations in the
B/C configuration at 1.4 GHz (\cite{ferra06}).
This relic source has been studied also at lower frequency with GMRT data
(\cite{gia06}, \cite{gia08}).

\begin{figure}
\centering
\includegraphics[width=9cm]{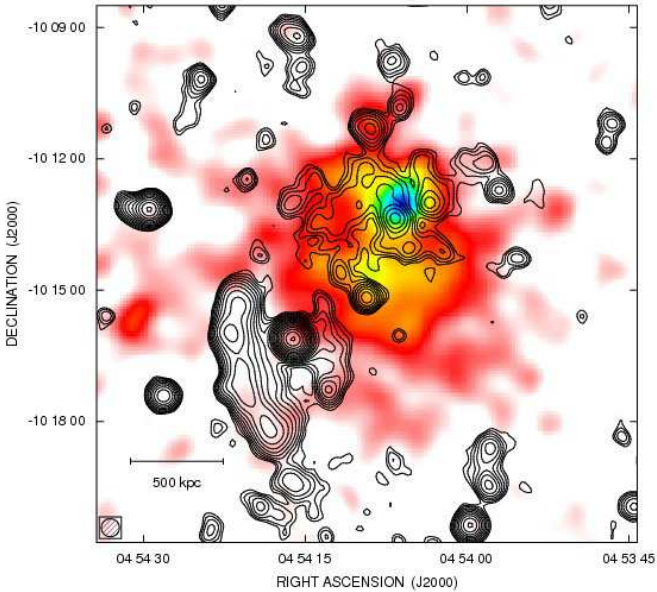}
\caption{Radio contours of the complex region in A521 obtained 
with the VLA at 1.4 GHz in B/C configuration.
The HPBW is $25'' \times 25''$ and the noise level is 0.03 mJy/beam. 
The central diffuse
halo source is well visible as well as the extended relic (see text).
The first contour level is drawn at 0.06 mJy/beam and the others
are spaced by a factor $\sqrt2$. The contours of the radio intensity 
are overlaid onto the Rosat HRI X-ray image in the 0.1-2.4 keV band. The 
X-ray image has been smoothed with a Gaussian of $\sigma=16''$.
}
\label{a521} 
\end{figure}

\cite{bru08} reported the detection of a diffuse halo at the cluster
center visible in GMRT data at 240, 325, and 610 MHz. They also reported a flux
density upper limit at 74 MHz from the VLA Low-frequency Sky Survey data
and discussed the missing detection at 1.4 GHz from VLA archive data.
They found that this halo has an extremely steep radio spectrum ($\sim$ 2.1)
with a high frequency cut-off.

We independently reanalyzed the VLA archive data to search
for a possible diffuse radio halo.
After carefully editing of the uv data we produced an image 
cutting the long baselines, and giving
a large weight to short baselines using ROBUST = 5 in the AIPS task IMAGR
(Natural weight).
The image  obtained 
clearly shows an extended emission not visible in previous images because of 
its low
brightness.  In Fig. \ref{a521} we present the radio contours of this 
low resolution image  overlaid onto the Rosat HRI X-ray image in 
the 0.1-2.4 keV band. The radio halo is 
well visible, and also the relic source
appears more diffuse than in previous images. Moreover a new extended 
emission almost parallel to the relic but less extended is visible.
The radio halo morphology in the present image is in good agreement with lower 
frequency images reported by \cite{bru08}, although their 
image at 240 MHz shows a larger extension because of a better uv-coverage 
and a larger beam, and with a more recent VLA image by \cite{dal09}.

We subtracted the contribution of discrete sources as discussed in 
Sect. 2.
The flux density emission from the diffuse source was S$_{tot}$ = 7.0 mJy, 
in the original image including discrete sources. After subtraction of 
discrete sources (1.1 mJy in total), the halo flux density is 5.9 mJy.

We compared our flux density measure with the values reported by \cite{bru08}:
(S$_{240}$ = 152 mJy, S$_{325}$ = 90 mJy, S$_{610}$ = 15 mJy), and derived 
a straight radio spectrum between 240, 325 and 1415 MHz with a constant 
spectral index $\alpha$ = 1.8. 
We note that the value at 610 MHz is lower than that implied by a straight 
spectrum, probably because of missing flux in GMRT data at this frequency. \\

\noindent
{\bf Abell 697}, a massive cluster at z = 0.282, shows
a central cD galaxy with a peculiar
disturbed morphology, indicative of a recent merger.  
The mass distribution is elongated in NE - SW direction. 
The central cD galaxy has a
double nucleus, which also supports the merger hypothesis. An
X-ray ROSAT HRI observation shows an elliptical
X-ray luminosity distribution in N-S direction (see also \cite{gir06}).

\begin{figure}
\centering
\includegraphics[width=9cm]{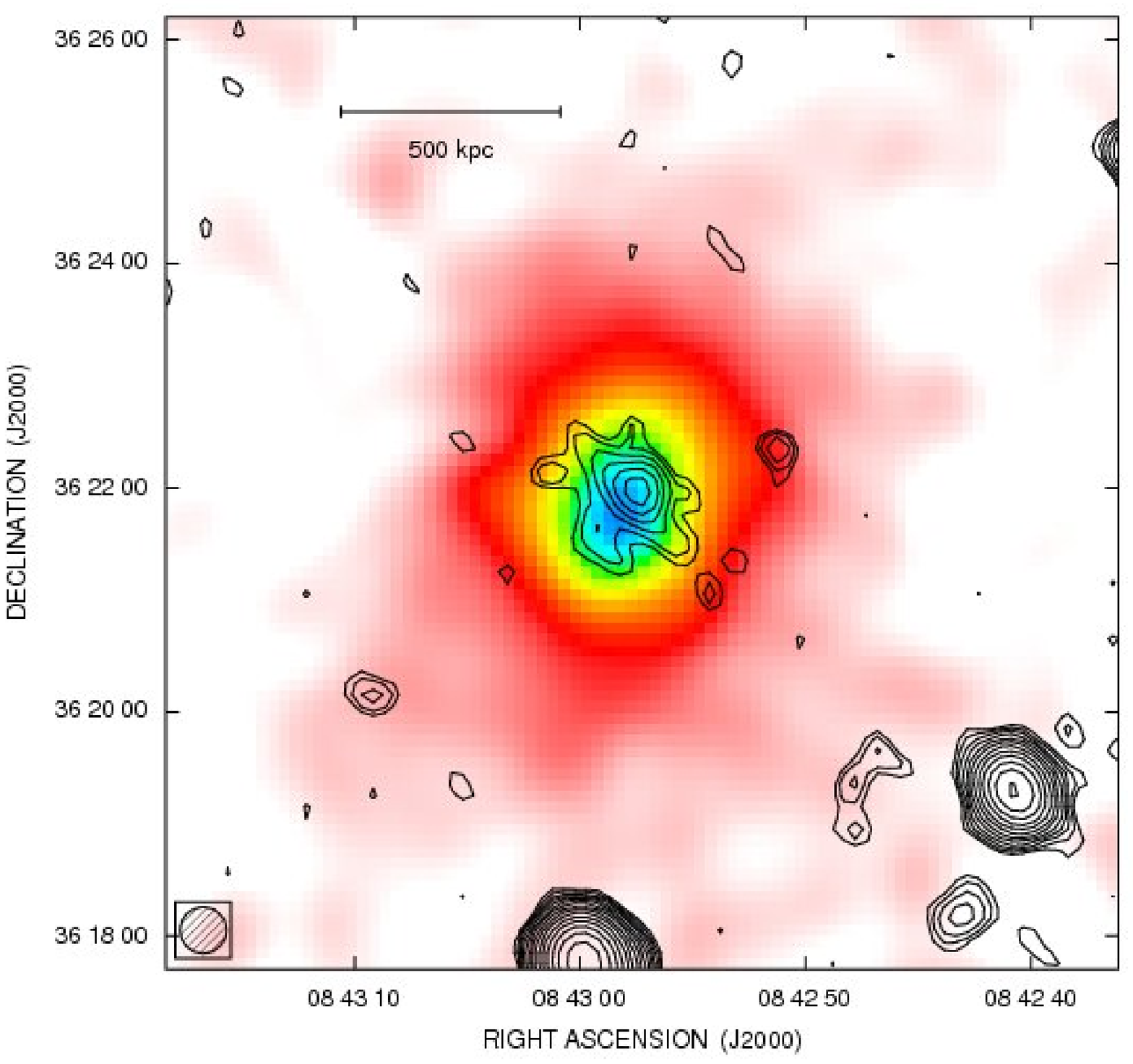}
\caption{Radio contours of A697 obtained with the VLA at 1.4 GHz 
in C configuration.
The HPBW is $25''\times 25''$ and the noise level is 0.04 mJy/beam. 
The first contour level is drawn at 0.1 mJy/beam and the others
are spaced by a factor $\sqrt2$. The contours of the radio intensity 
are overlaid onto the Rosat HRI X-ray image in the 0.1-2.4 keV band. The 
X-ray image has been smoothed with a Gaussian of $\sigma=16''$.}
\label{a697} 
\end{figure}

In the radio band an extended halo was detected by \cite{ven08} using 
GMRT data at 610 MHz and recently at 325 MHz (\cite{ven09}). 
The halo flux density at 325 MHz is 45 mJy and 13 mJy at 610 MHz, resulting
in a very steep radio spectrum ($\alpha^{0.6}_{0.3} \sim$ 2, \cite{ven09}).
Here we present a new image obtained using archive
VLA data at 1415 MHz in the C configuration. 
In Fig. \ref{a697} we present the radio contours of this 
image  overlaid onto the Rosat HRI X-ray image in 
the 0.1-2.4 keV band. The image confirms the
presence of central diffuse emission. The total flux density is 1.7 mJy,
and the noise level is 0.1 mJy/beam.
A comparison between present data, the NVSS, and \cite{ven08}
shows that in the C array VLA data most of the
extended flux is missing (indeed the NVSS flux density is 7.8 mJy).
Therefore we report in Tab. 2 
the 1.4 GHz flux density
and size estimated from the NVSS.
Comparing NVSS and GMRT data at 325 MHz we derive a halo spectral index 
$\alpha^{1.4}_{0.3} \sim$ 1.2.
A deeper VLA image at 1.4 GHz is necessary to 
confirm this result. \\

\begin{figure*}[!!htb]
\centering
\includegraphics[width=18cm]{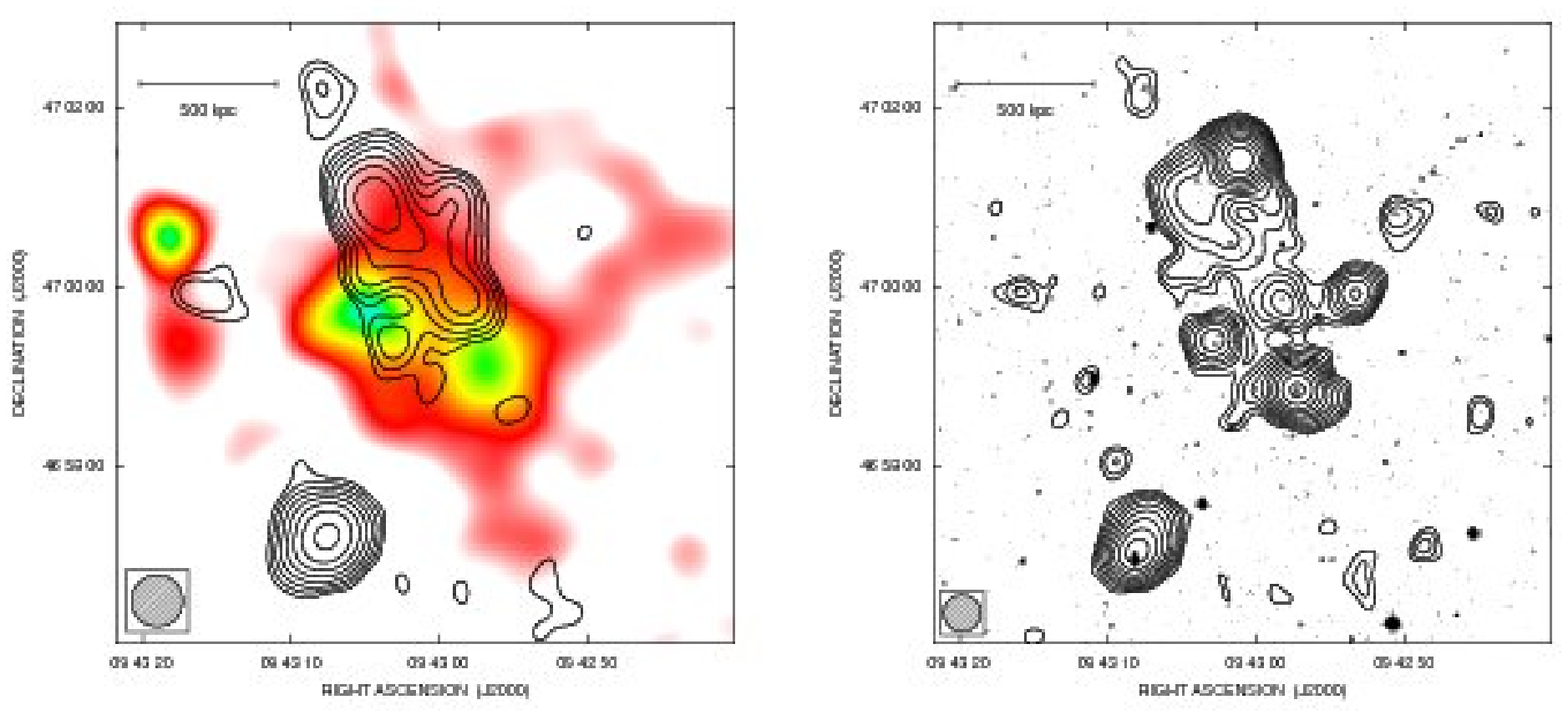}
\caption{Left: radio contours of A851 obtained with the VLA at 1.4 GHz 
combining data in C and D configuration, discrete sources in the cluster
region have been subtracted.
The HPBW is $35'' \times 35''$ and the noise level is 0.03 mJy/beam.
The first contour level is drawn at 0.09 mJy/beam and the others
are spaced by a factor $\sqrt2$. The contours of the radio intensity
are overlaid onto the Rosat HRI X-ray image in the 0.1-2.4 keV band. 
The X-ray image has been smoothed with a Gaussian of $\sigma=16''$.
Right: radio contours as in the left figure but with no source subtraction.
The HPBW is $25'' \times 25''$.  The contours of the radio intensity are
overlaid onto the optical image from the POSS2 red plate.} 
\label{a851} 
\end{figure*}

\begin{figure}[!!htb]
\centering
\includegraphics[width=9cm]{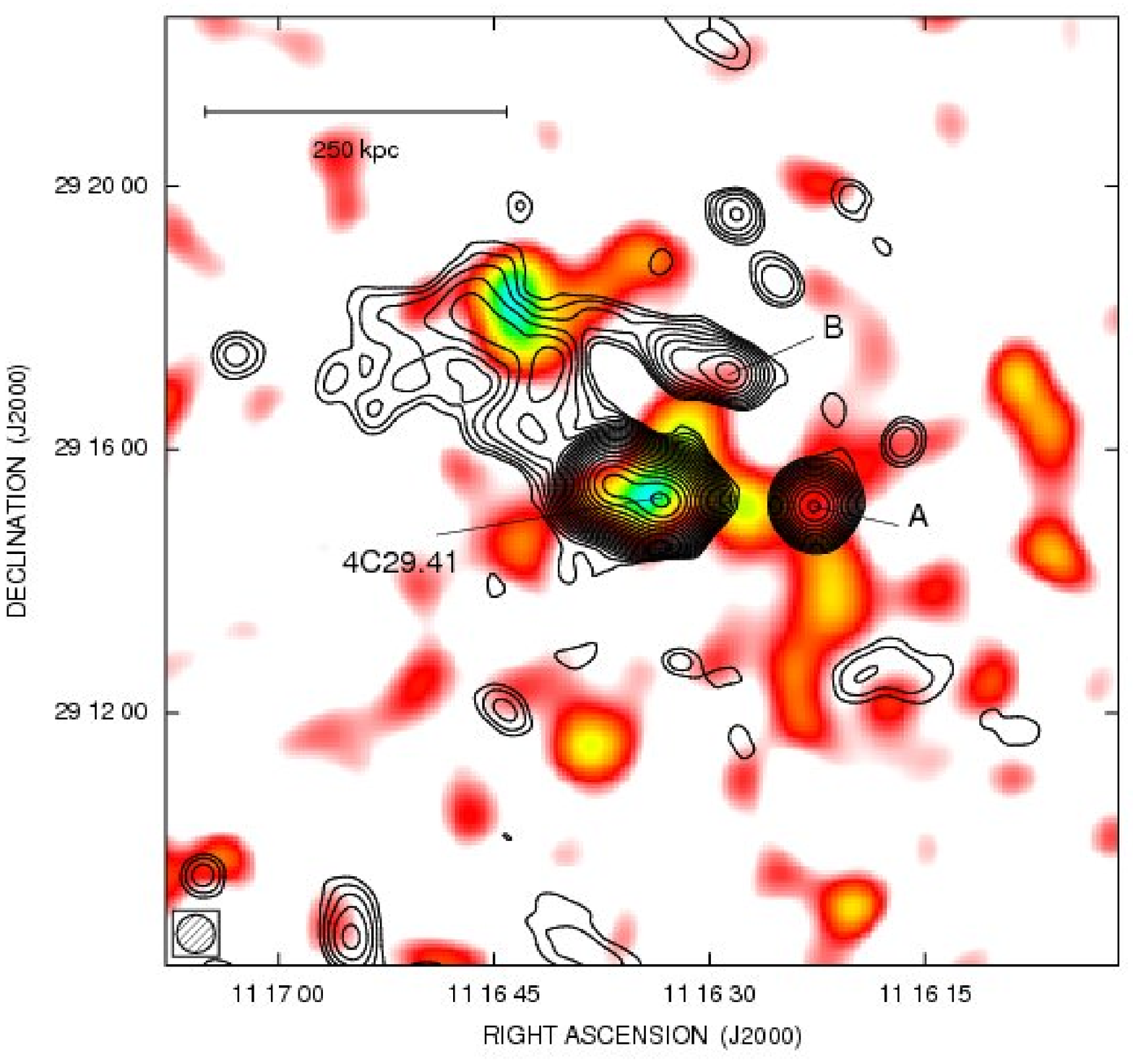}
\caption{Radio contours of A1213  obtained 
with the VLA at 1.4 GHz in C+C/D configuration.
The HPBW is $35'' \times 35''$ and the noise level is 0.3 mJy/beam. 
The first contour level is drawn at 1 mJy/beam and the others
are spaced by a factor $\sqrt2$. The contours of the radio intensity 
are overlaid onto the Rosat HRI X-ray image in the 0.1-2.4 keV band. The 
X-ray image has been smoothed with a Gaussian of $\sigma=16''$.
Cluster radio galaxies (the BCG 4C29.41, A, and B) are named according to the 
text.}
\label{a1213} 
\end{figure}

\noindent
{\bf Abell 851 - CL0939+4713} is a rich cluster of galaxies at z = 0.4069.
A study by \cite{sm06} shows a high number of starburst
galaxies probably due to
some cluster-specific mechanism, likely related to the dynamical
assembly of the cluster. The existence of an extended diffuse emission in this
cluster was first reported by \cite{owe99}.
We reduced C, and D archive VLA data at 1400 MHz.
In Fig. \ref{a851} we present the radio contours of A851 
 overlaid onto the optical image from the POSS2 red plate (right) 
and radio contours
of the diffuse emission (after subtracion of discrete sources) overlaid onto
the Rosat HRI X-ray image in 
the 0.1-2.4 keV band (left). 
We note that the X-ray image shows a double structure
separated by less than 2'. The radio diffuse emission is asymmetric, on the 
Northern side with respect to the X-ray brightness peaks. 
The estimated total flux
density of the extended radio complex is 11.9 mJy. After subtraction of 
discrete sources (8.2 mJy),
the halo total flux is 3.7 mJy. \\

\noindent
{\bf Abell 1213} is a poor nearby Abell cluster at z = 0.0469,
characterized by the 
presence of a dumbell radio galaxy (the BCG) identified with a double radio 
source (4C29.41, B2 1113+29, \cite{fan82}). 
In the X-band it was detected by the 
Einstein 
(\cite{jf99}) and Rosat (\cite{led03}) satellites. The X-ray 
image shows a faint irregular emission. The estimated X-ray luminosity
in the 0.1 - 2.4 keV band is 0.10 $\times$ 10$^{44}$ erg/sec i.e. at least 10 
times lower than that of any cluster presently known to host a radio halo 
(see Table 2).
Optical and X-ray data suggest
a non-relaxed structure.

\begin{figure*}[!!htb]
\centering
\includegraphics[width=18cm]{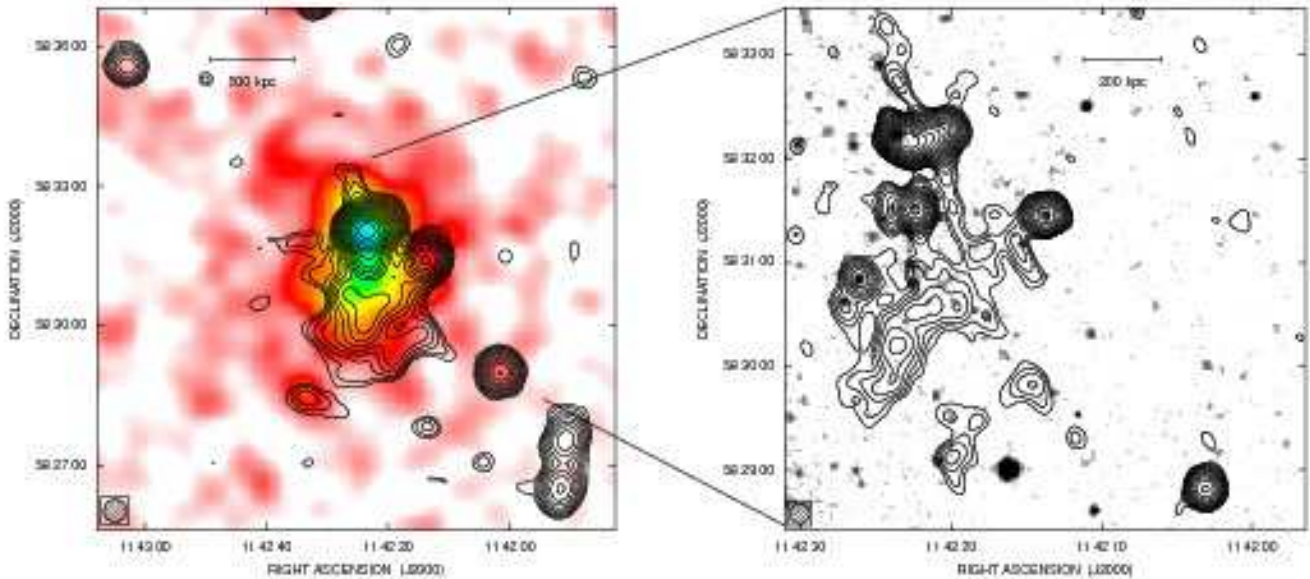}
\caption{
Left: radio contours of the extended halo in A1351 obtained with the VLA 
at 1.4 GHz combining data in C+D
 configuration.
The HPBW is $30'' \times 30''$ and the noise level is 0.09 mJy/beam.
The first contour level is drawn at 0.25 mJy/beam and the others
are spaced by a factor $\sqrt2$. The contours of the radio intensity
are overlaid onto the Rosat HRI X-ray image in the 0.1-2.4 keV band. The X-ray 
image has been smoothed with a Gaussian of $\sigma=16''$.
Right: radio contours obtained with the VLA at 1.4 GHz in C 
configuration of A1351.
The HPBW of the radio image is $11'' \times 11''$ and the noise level is 
0.06 mJy/beam. The first contour level is drawn at 0.15 mJy/beam and the others
are spaced by a factor $\sqrt2$. The contours of the radio intensity are 
overlaid onto the optical image from the POSS2 red plate.}
\label{a1351} 
\end{figure*}

The radio morphology of the BCG is peculiar being a small-size bright 
double similar to FR II radio galaxies. Moreover despite the small number of
bright galaxies, at least two other cluster members are identified as 
radio galaxies (A, B see \cite{fan82} and Fig. \ref{a1213}). 
The FIRST survey (\cite{bec95}) confirms \cite{fan82} results,
while the NVSS detected a diffuse extended emission which cannot be due to
the presence of the discrete sources found in the FIRST image. 
The extended emission 
is off-center with respect to the BCG (NE
direction). To better investigate this structure we obtained short 
VLA observations pointed on the cluster center in the C and C/D configuration.
Our final image is presented in Fig. \ref{a1213}. The extended diffuse emission
is easily visible. It is similar to small-size asymmetric radio halos 
found in other clusters as A2218 (\cite{gf00}), and we 
classify it as a small-size radio halo.
However the diffuse radio source shows a peculiar bright filament in 
the inner region with a
sharp bend as a Z shaped structure. This filament is in direction of the BCG
bright radio galaxy but it is not connected to it. 
The radio structure
of 4C29.41 and the size of the diffuse component excludes a possible 
connection between the BCG and the diffuse structure due to 
a possible proper motion of the BCG. 

A possible connection is also
suggested by the radio morphology of the diffuse emission, with the source at 
north of the BCG, identified with a 
cluster elliptical galaxy (source B). 
This source shows a peak flux density in the FIRST images = 5.3 mJy/beam.
It appears slightly resolved on the W side with respect to the galaxy position,
i.e. on the opposite side to the extended diffuse emission. We cannot exclude
that this source could be a head-tail  radio galaxy, but the 
radio morphology and radio power rule out the possibility that all or most
of the diffuse emission could be identified as an extended tailed structure.
Therefore as discussed before we connect the extended source not to the
activity of  one or more 
cluster galaxies but to the physical properties of the whole cluster .\\ 

\noindent
{\bf Abell 1351} is a rich massive cluster at z = 0.3224 with an elongated 
X-ray brightness distribution indicating a possible ongoing merger.
It shows also an unusually high velocity dispersion, and the presence
of a bright red gravitational arc offset from the cluster light
center (\cite{dah02}).

\cite{owe99} reported the existence of a halo source in this
cluster. We analyzed VLA archive data at 1.4 GHz in C and D configuration.

In Fig. \ref{a1351} we show the cluster radio emission
compared with optical and X-ray images.
We produced a low resolution radio image to obtain a better signal 
to noise ratio
for the halo source. On the left the contours of this image
are overlaid 
onto the Rosat HRI X-ray image in the 0.1-2.4 keV band.
On the right we show a zoom of radio contours at higher resolution  
 overlaid 
onto an optical image.
The presence of
an extended diffuse radio halo source is easily visible even in the 
high resolution 
image which  also shows 4 discrete sources. Two of these sources show
an elongated head-tail like structure and are identified as possible
cluster galaxies. We note that the radio-diffuse 
emission is centrally located but asymmetric with respect to the X-ray 
brightness peak.

In this cluster the subtraction of discrete sources was
more problematic, because of the presence of two strong, extended 
tailed radio galaxies
in the Northern region (see Fig. 6, right). For this reason we present
in Fig. 6 left the image with no source subraction. We checked 
the total flux density of the diffuse 
emission, obtained as discussed in Sect. 2, with the halo 
flux density obtained by integrating the central cluster region in
the low resolution image, and subtracting the estimated flux density of
discrete sources derived from a multi-component gaussian fit 
of discrete sources (JMFIT in AIPS). 
We found a good agreement between the two values.\\ 

\noindent
{\bf Abell 1758} from the optical point of view, is a double cluster
which is possibly in the process of merging into a single massive
cluster. Both sub-clumps are
concentrated around a bright early-type galaxy. There is an
apparent blue arc associated with the northwest mass clump (\cite{dah02}).

It was studied in detail in the X band
by \cite{dk04}. They
confirm the double structure of this cluster: it consists of two hot X-ray 
luminous clusters:
A1758N and A1758S. Moreover the two main clusters show a clear substructure:
A1758N is in the late
stage of a merger of two 7 keV subclusters, while A1758S is in
the early stages of a merger of two 5 keV subclusters.

\begin{figure}[!!htb]
\centering
\includegraphics[width=9cm]{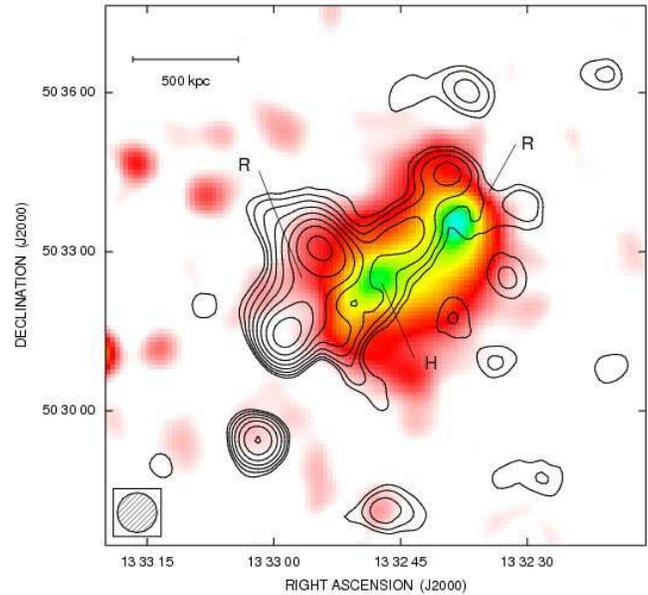}
\caption{Radio contours of A1758N obtained with the VLA at 1.4 GHz 
in configuration D, after subtraction of discrete sources.
The HPBW is $45'' \times 45''$ and the noise level is 0.07 mJy/beam.
The first contour level is drawn at 0.2 mJy/beam and the others
are spaced by a factor $\sqrt2$. H and R show the halo and relic 
sources, see text.
The contours of the radio intensity
are overlaid onto the Rosat HRI X-ray image in the 0.1-2.4 keV band. 
The X-ray image has been smoothed with a Gaussian of $\sigma=16''$.
}
\label{a1758} 
\end{figure}

\cite{gio06} reported the detection of a radio halo in A1758N. 
Radio images show that the central emission is confused by unrelated discrete 
sources. After subtraction in the uv-plane we obtained a low resolution image
of the extended emission.
In Fig. \ref{a1758} we present the radio contours of the extended halo 
source with the discrete sources subtracted.
For a comparison between the radio and the X-ray cluster emission,  
the contours of the radio intensity are overlaid 
onto the Rosat HRI X-ray image in the 0.1-2.4 keV band of A1758N.
The two peaks corresponding to the subclusters in the late
merging phase are clearly visible.
The radio image shows a complex structure similar to that of A754 
(\cite{kas01}, \cite{bac03}). In A1758N we identify
a central diffuse emission (halo) permeating the
central region where the two sub-clusters are merging, 0.8
Mpc in size (H in Fig. \ref{a1758}), 
and two brighter peripheral structures 
on the opposite
side with respect to the cluster center, which resemble
relic radio sources R in Fig. \ref{a1758}). 
A1758N would be one of the very few clusters with a 
central halo and two peripheral relics as in RXCJ1314.4-2515 
(\cite{fer05}). 

The detection of a radio-diffuse emission
in A1758N and not in A1758S is consistent with the hotter
temperature (7keV) of the Northern subclusters with respect to
Southern subclusters (5 keV) and with the results of \cite{dk04}:
that A1758N is in a late stage of merger, while A1758S is in
an early merger stage.

\skuno

\noindent
{\bf Abell 1995} is a rich massive cluster at z = 0.3186.
The light distribution in this cluster is quite strongly
elongated in the northeast-southwest direction, but the mass
distribution is more circularly symmetric and
concentrated. Moreover, \cite{dah02} discussed the presence
of several blue arcs.
The X-ray morphology
analyzed by \cite{pat00}
appears very similar to the mass distribution.

\begin{figure*}[!!htb]
\centering
\includegraphics[width=18cm]{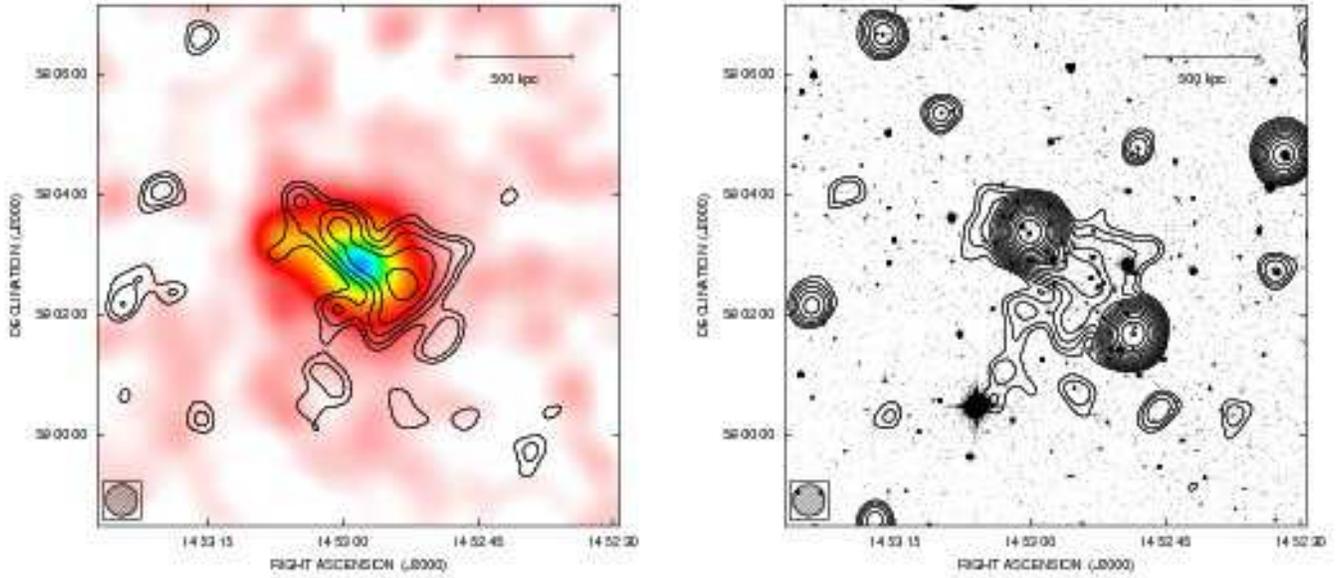}
\caption{Left: Radio contours of A1995  obtained 
with the VLA at 1.4 GHz in C+D configuration, after subtraction of discrete
sources.
The HPBW is $30'' \times 30''$ and the noise level is 0.05 mJy/beam. 
The first contour level is drawn at 0.1 mJy/beam and the others
are spaced by a factor $\sqrt2$. The contours of the radio intensity 
are overlaid onto the Rosat HRI X-ray image in the 0.1-2.4 keV band. The 
X-ray image has been smoothed with a Gaussian of $\sigma=16''$. Right: VLA
radio image as in 
the left, but with no source subtraction. The radio contours are overlaid 
onto the optical image from the POSS2 red plate.}
\label{a1995} 
\end{figure*}

\cite{owe99} reported the existence of a halo source in this
cluster. We reduced VLA archive data in C and D configuration. An extended 
diffuse emission is clearly visible at the cluster center (Fig. \ref{a1995} 
left). 
Two relatively
bright point-like sources (Fig. \ref{a1995} right) 
have been subtracted to estimate the halo parameters.
The radio contours of the halo sources after subtraction of discrete sources 
are overlaid onto the Rosat HRI X-ray image in the 0.1-2.4 keV band. 
The original VLA image is overlaid
onto the optical image from the POSS2 red plate. \\

\noindent
{\bf Abell 2034} shows multiple 
signatures of an ongoing merger in a Chandra X-ray image, 
including a cold front and probable 
significant heating of the ICM (\cite{kem03}). In this region \cite{ks01} 
detected an extended radio emission North of the cluster
center near to the position of the cold front. We analyzed archive VLA
data at 1.4 GHz in C and D configuration to confirm and to image in
detail the diffuse emission.

\begin{figure*}[!!htb]
\centering
\includegraphics[width=18cm]{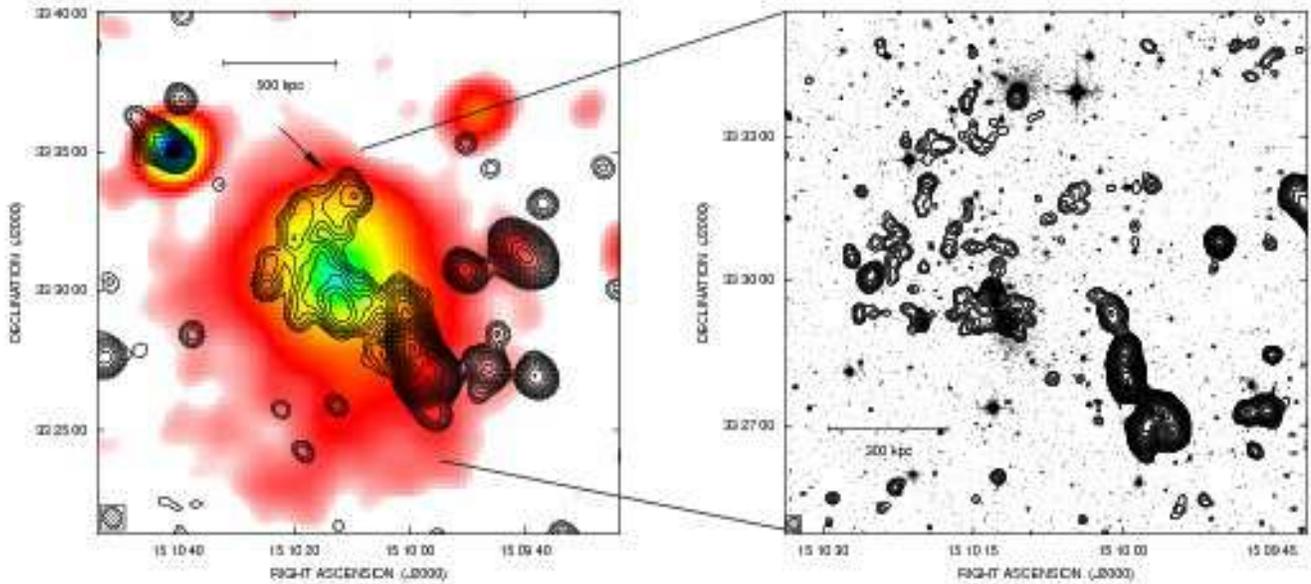}
\caption{
Left: radio contours of the halo in A2034 obtained with the VLA 
at 1.4 GHz combining data in C+D configuration.
The HPBW is $43.8'' \times 39.9''$ in PA 33$^\circ$ 
and the noise level is 0.04 mJy/beam.
The first contour level is drawn at 0.1 mJy/beam and the others
are spaced by a factor $\sqrt2$. The contours of the radio intensity
are overlaid onto the Rosat PSPC X-ray image in the 0.1-2.4 keV band. 
The X-ray image has been smoothed with a Gaussian of $\sigma=30''$.
Right: radio contours obtained with the VLA at 1.4 GHz in C configuration 
of A2034.
The HPBW of the radio image is $15.7'' \times 14.9''$ in PA -3$^\circ$
and the noise level is 
0.016 mJy/beam. The first contour level is drawn at 0.04 mJy/beam 
and the others
are spaced by a factor $\sqrt2$. The contours of the radio intensity are 
overlaid onto the optical image from the POSS2 red plate. 
An arrow points the cold front 
region (see text)
}
\label{a2034} 
\end{figure*}

In Fig. \ref{a2034} we show the cluster radio emission
compared with the optical and X-ray images.
On the left the contours of a low resolution image 
are overlaid 
onto the Rosat PSPC X-ray image in the 0.1-2.4 keV band.
In the right panel we show a zoom of radio contours at higher resolution  
 overlaid onto the optical image taken from 
the POSS2 red plate in the 
Optical Digitized Sky Survey.
The diffuse radio emission is located at the cluster center, but with respect
to other giant halos it appears elongated and irregular. The Northern region
(pointed out by an arrow)
of the diffuse source coincides with the tentative relic in the cold front
region discussed by \cite{kem03}.
\\

\begin{figure*}[!!htb]
\centering
\includegraphics[width=18cm]{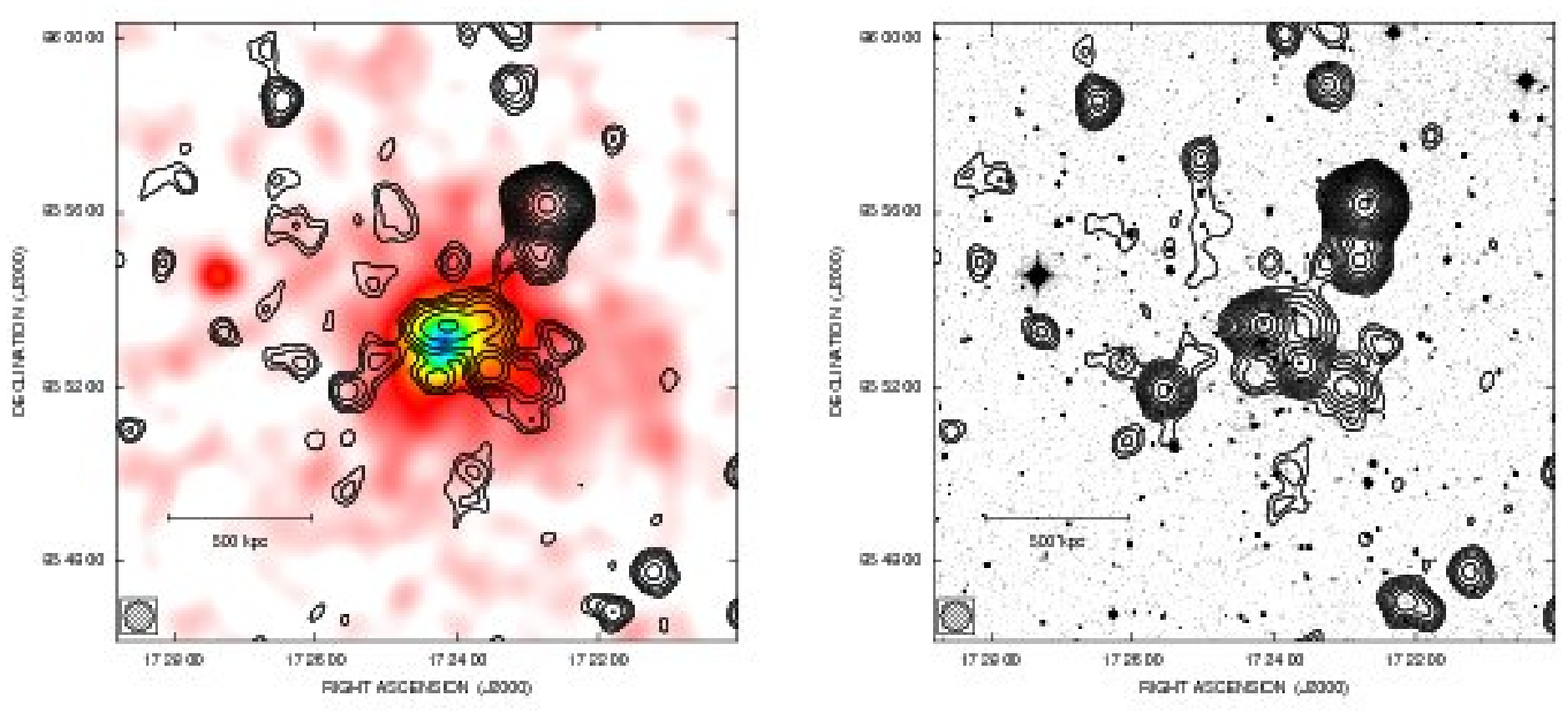}
\caption{
Left: radio contours of A2294 obtained with the VLA at 1.4 GHz 
combining data in C and D configuration. Discrete sources in the cluster
region have been subtracted.
The HPBW is $35'' \times 35''$ and the noise level is 0.04 mJy/beam.
The first contour level is drawn at 0.12 mJy/beam and the others
are spaced by a factor $\sqrt2$. The contours of the radio intensity
are overlaid onto the Rosat HRI X-ray image in the 0.1-2.4 keV band. 
The X-ray image has been smoothed with a Gaussian of $\sigma=16''$.
Right: as in the Left panel but without source subtraction.  
The contours of the radio intensity are
overlaid onto the optical image from the POSS2 red plate.}
\label{a2294} 
\end{figure*}

\noindent
{\bf Abell 2294} is a rich massive cluster at z = 0.178 
the temperature profile is consistent with an isothermal
cluster at T $\sim$ 8 keV (\cite{riz98}).
\cite{owe99} reported the existence of a halo source in this
cluster. We analyzed VLA data in the C and D configuration. 
In Fig. \ref{a2294} we present 
the radio contours of A2294 after subtraction
of discrete sources, overlaid onto the Rosat HRI X-ray image in 
the 0.1-2.4 keV band (left), and without source subtraction  
on the optical image from the POSS2 red plate (right).
The diffuse radio emission, 3' in size, is in good 
agreement with the cluster X-Ray image. \\
 
\noindent 
{\bf Abell 3444} is a rich cluster at z = 0.2533.
\cite{mat01} were able to fit the ASCA spectra 
with a single-temperature thermal plasma model (5.7 keV).
\cite{lem99} and \cite{mat01}) defined 
A3444 as a possible cooling core cluster.
Detailed optical studies are not available for this cluster. 

We reduced VLA archive data at 1.4 GHz in A/B and C/D configuration.
The low resolution image shows a very extended faint radio filament,
located around a chain
of discrete radio sources. 
\cite{ven07} observed this cluster with the GMRT at 610 MHz and found that
discrete radio sources have an optical counterpart and
follow the inner elongation of the archive ASCA X-ray image.
Moreover in GMRT images the dominant cluster galaxy shows 
an extended radio emission, and a flux density of 10
mJy is present on a scale of 1.5' at 610 MHz.

\begin{figure*}[!!htb]
\centering
\includegraphics[width=18cm]{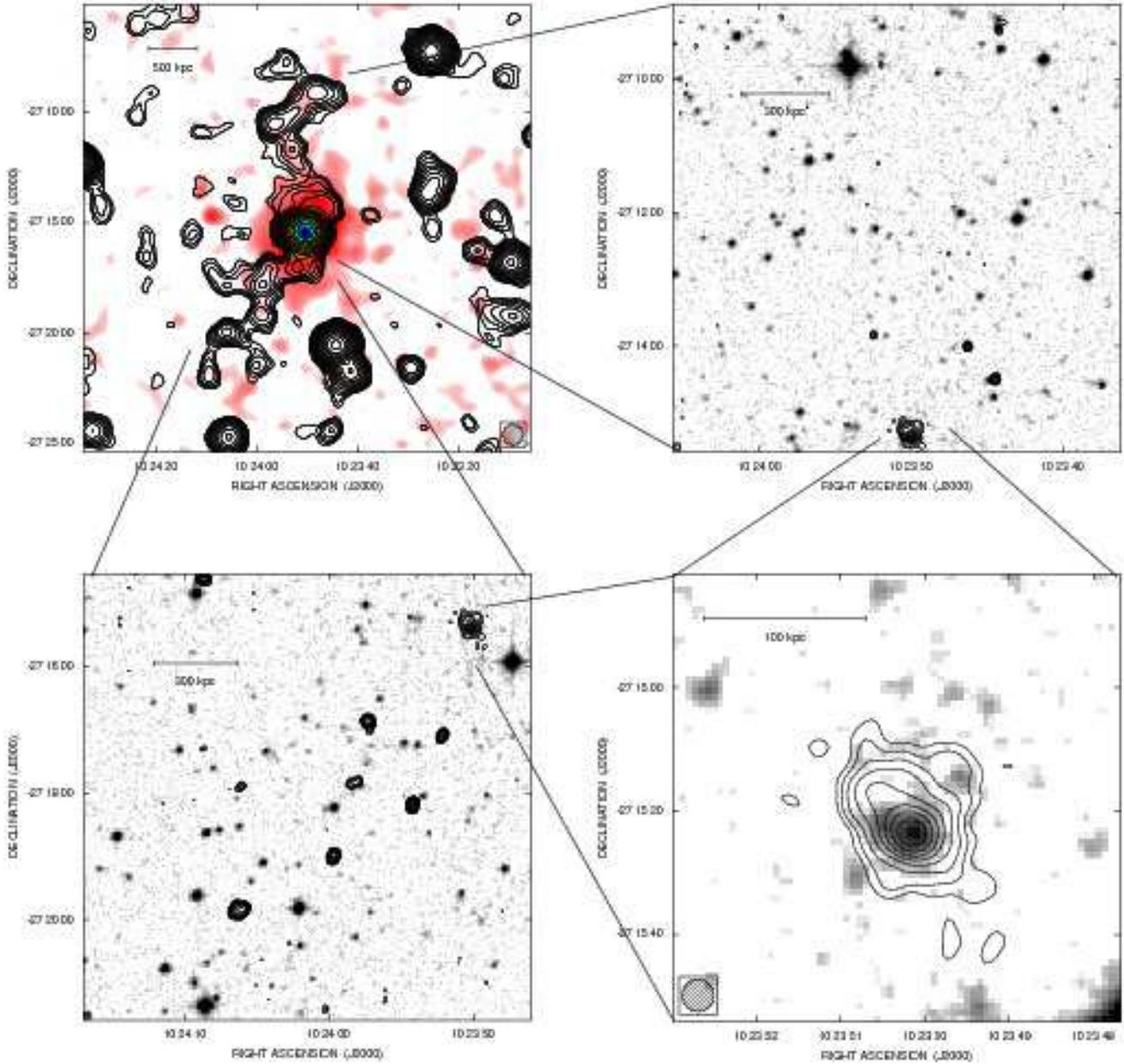}
\caption{
Top, Left: radio contours of the diffuse filament located in A3444 
obtained with the VLA at 1.4 GHz combining data in A/B and C/D configuration,
after subtraction of discrete sources (but not the central cD).
The HPBW is $55'' \times 55''$ and the noise level is 0.033 mJy/beam.
The first contour level is drawn at 0.1 mJy/beam and the others
are spaced by a factor $\sqrt2$. The contours of the radio intensity
are overlaid onto the Rosat HRI X-ray image in the 0.1-2.4 keV band. 
The X-ray image has been smoothed with a Gaussian of $\sigma=16''$.
In this figure the restoring beam is in the bottom right corner.
Top, Right: zoom to the north of the central cD galaxy of the VLA high 
resolution radio image. 
The HPBW is 5$'' \times 5''$ and the noise level is 0.03 mJy/beam. 
The first contour level is drawn at 0.1 mJy/beam and the others
are spaced by a factor $\sqrt2$. The contours of the radio intensity are 
overlaid onto the red optical image from the SERC/ESO surveys.
Bottom, left: zoom to south of the central cD galaxy. 
Bottom, right: zoom of the central cD in A3444.
}
\label{a3444} 
\end{figure*}

Our high resolution image
is presented on the top right and bottom panels of Fig. \ref{a3444},
where zooms of the radio contours in different regions of the radio
filament are  overlaid onto an optical image. 
The radio emission of the central cD galaxy is
clearly extended with a size of 45$''$. The peak is 6.5 mJy/beam,
and the total flux is 7.4 mJy. We classify this extended emission as a 
faint 'mini-halo' diffuse source in agreement with the presence of
a possible central cooling flow.

We subtracted in the uv-plane all the discrete sources embedded in the 
extended emission except the central cD galaxy, since it is clearly resolved
in our images, so as
to study a possible connection between it and the extended emission. 
This image 
is presented in Fig. \ref{a3444} where the radio contours are overlaid 
onto the Rosat HRI X-ray image in the 0.1-2.4 keV band.

A clear filamentary 
structure $\sim$ 14' in size corresponding to 3.3 Mpc at the cluster distance
is present, centered on the cD galaxy.
However because of the large size and morphology we exclude a direct connection
between the cD and this radio filament. We do not classify this extended and
diffuse radio emission as a radio halo, but we suggest a connection
with a possible filamentary structure as found by \cite{bag02},
and \cite{br09}.
We note that the presence of a giant filamentary radio structure
is not incompatible with a possible central cooling flow in A3444. 
\\

\begin{figure}[!!htb]
\centering
\includegraphics[width=9cm]{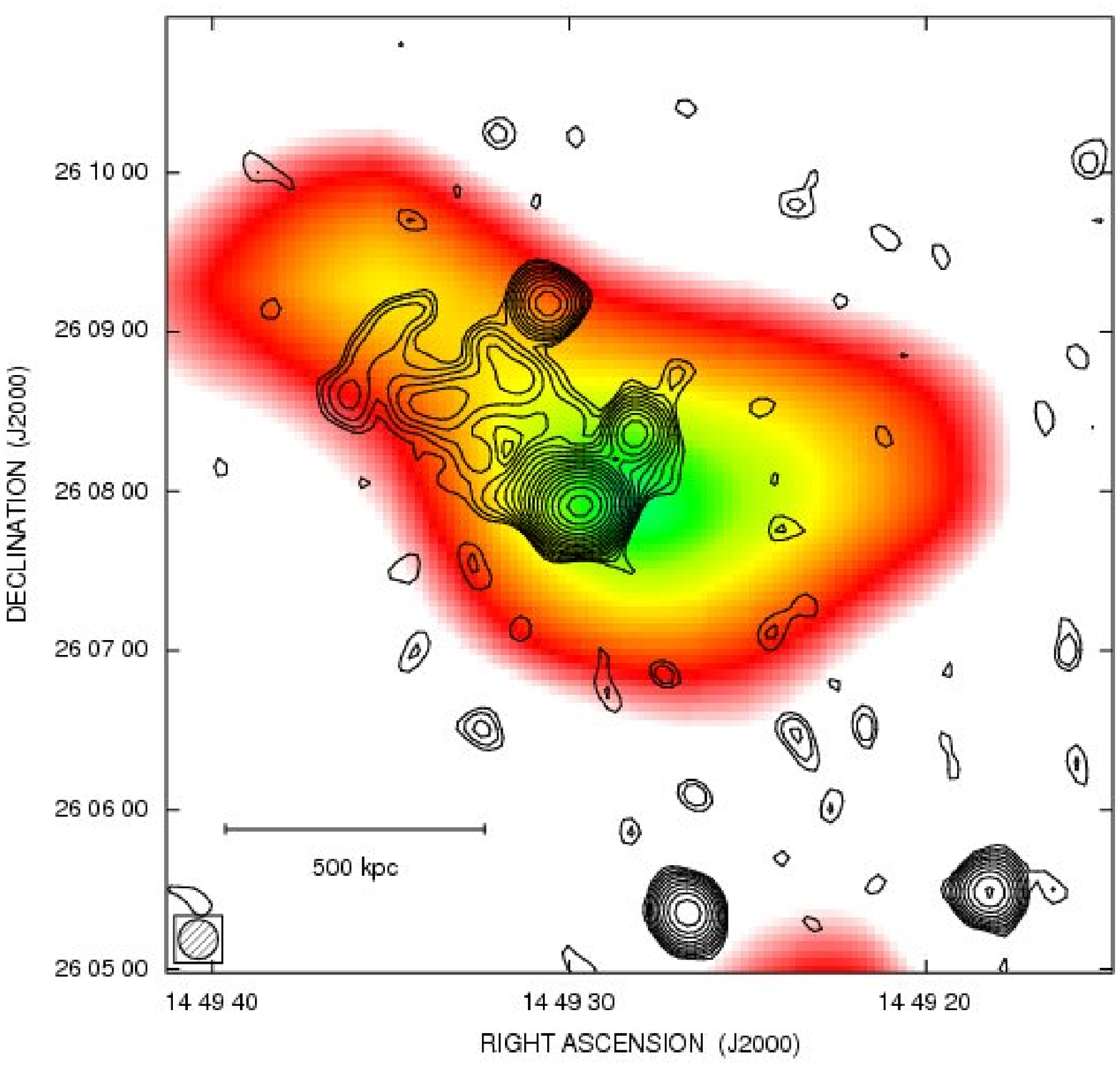}
\caption{
Radio contours of the central region in CL1446+26  
obtained with the VLA at 1.4 GHz in C configuration.
The HPBW is $15'' \times 15''$ and the noise level is 0.05 mJy/beam.
The first contour level is drawn at 0.1 mJy/beam and the others
are spaced by a factor $\sqrt2$. The contours of the radio intensity
are overlaid onto the Rosat All Sky Survey image in the 0.1-2.4 keV band. 
The X-ray image has been smoothed with a Gaussian of $\sigma=45''$.}
\label{cl1447} 
\end{figure}

\noindent
{\bf CL 1446+26} named also CL 1447+26 or ZwCl 1447.2+2619,  
has been discussed 
by \cite{owe99} who show a radio image 
overlayed onto an optical R-band image.
We retrieved an archive VLA observation in C configuration and obtained a 
deeper image shown in Fig.\ref{cl1447}.
The radio contours are
 overlaid onto the Rosat All Sky Survey X-ray image in 
the 0.1-2.4 keV band.
 
From a comparison with the FIRST and the \cite{owe99} image,
it is clear that in this cluster two head-tail radio galaxies are present
and appear blended in a strong unresolved source in our image at RA=14 49 30, 
DEC=$+$26 08 00 in the periphery of the diffuse emission.
A third head tail radio galaxy is at RA=14 49 28 DEC=$+$26 08 20. 
Moreover a background source is north of the extended emission
at RA=14 49 30 DEC=$+$26 09 10. 

The X-Ray image shows an elongated emission with the presence of a 
secondary peak displaced by about 500 kpc up to the NE. We interpret this 
substructure as evidence of a smaller group merging into the main cluster.
The diffuse radio emission is in-between the two X-Ray peaks and for this 
reason  it is more appropriate to identify it as a relic radio 
source. 

\section{Discussion}
Next we present a few statistical considerations on the properties of 
radio halos combining radio halos taken from the literature with those
presented above.
We note that this is not a complete sample of radio halos therefore we
cannot use it to analyze statistical properties such as the percentage
of clusters with a radio halo or the radio halo luminosity function.
However, the large number of halo sources collected here
can be used to derive correlations and to obtain the observational 
properties of known radio halos.

In Tab. 2 we present all the known halos at z$<$ 0.4
with a high quality pointed radio image, including literature data and
sources presented for the first time in this work
(see col. 8 for the references related to the radio data). 
Radio halos at larger
redshift are rare. We will present results for higher
redshift clusters in a forthcoming paper (but see \cite{gf00}, and
\cite{bon09b}). 
We have a total of 33 sources;
however, we note that in the case of CL1446+26 the diffuse source should be
classified as a relic source and in A3444 the giant diffuse emission 
is likely a giant filament, and the extended emission around the
cD galaxy a mini-halo. The final number of certain radio halos
is therefore 31.
The flux density of radio halos at 1.4 GHz and the estimated error 
are indicated in col. 4 and col. 5 of Tab. 2. 
The radio power at 1.4 GHz, and the 
radio Largest Linear Size (LLS)
are given in col. 6 and col 7. 
For each cluster in col. 8 we indicate the X-ray luminosity in the 0.1-2.4 keV
band 
obtained from the references given in col. 10 and converted to the cosmology
used in this paper.
In the case of A1213 the X-ray emission in the 0.5-2 keV band 
(\cite{led03}) has been converted to the 0.1-2.4 keV band
using Tab. 4 of \cite{boe04},
while for CL1446+26 the bolometric X-ray emission (\cite{wu99}) has 
been converted to the 0.1-2.4 keV band using Tab. 5 of \cite{boe04}. 
In both cases we assumed an intra-cluster temperature value T$_X$ 
$\simeq$ 5 keV.

\begin{table*}[!!htb]
\caption{Radio halos}
\label{table2}
\centering
\begin{tabular}{lclrrrrrccc}
\hline\hline
Name    & z & kpc/$''$ & S(1.4) & $\Delta$S  &log P(1.4) & LLS & Lx(10$^{44}$)  & Ref. & Ref.  & notes \\
        &   &          & mJy    & mJy        & W/Hz      & Mpc & erg/sec        & Radio& X-ray &       \\
\hline
A209    & 0.2060 & 3.34 & 16.9 & 1.0 & 24.31 & 1.40 & 6.17 & * & 21 &a relic could be present \\
A401    & 0.0737 & 1.38 & 17.0 & 1.0 & 23.34 & 0.52 & 6.52 & 1 & 15 & elongated and irregular \\
A520    & 0.1990 & 3.25 & 34.4 & 1.5 & 24.58 & 1.11 & 8.30 & 2 & 16 & \\
A521    & 0.2533 & 3.91 &  5.9 & 0.5 & 24.05 & 1.17 & 8.47 & * & 21 & a relic is also present \\
A545    & 0.1540 & 2.64 & 23.0 & 1.0 & 24.16 & 0.89 & 5.55 & 1 & 21 & \\
A665    & 0.1819 & 3.03 & 43.1 & 2.2 & 24.59 & 1.82 & 9.65 & 3 & 16 & \\
A697    & 0.2820 & 4.23 &  7.8 & 1.0 & 24.28 & 0.65 &10.40 & * & 16 &from NVSS, see text \\
A754    & 0.0542 & 1.04 & 86.0 & 4.0 & 23.77 & 0.99 & 2.21 & 1 & 15 & a relic is also present \\
A773    & 0.2170 & 3.48 & 12.7 & 1.3 & 24.23 & 1.25 & 7.95 & 2 & 16 & \\
A851    & 0.4069 & 5.40 &  3.7 & 0.3 & 24.33 & 1.08 & 5.04 & * & 17 &   \\
A1213   & 0.0469 & 0.91 & 72.2 & 3.5 & 23.56 & 0.22 & 0.10 & * & 18 & asymmetric \\
A1300   & 0.3072 & 4.49 & 20.0 & 2.0 & 24.78 & 1.3  &13.73 &4,5& 21 & a relic is also present \\
A1351   & 0.3224 & 4.64 & 39.6 & 3.5 & 25.12 & 0.84 & 5.47 & * & 17 &         \\
A1656   & 0.0231 & 0.46 &530.0 &50.0 & 23.80 & 0.83 & 3.99 & 6 & 15 & a relic is also present \\ 
A1758   & 0.2790 & 4.20 & 16.7 & 0.8 & 24.60 & 1.51 & 7.09 & * & 19 &total diffuse emission H+R \\
        &        &      &  3.9 & 0.4 & 23.97 & 0.63 &      & * &    & central halo \\
A1914   & 0.1712 & 2.88 & 64.0 & 3.0 & 24.71 & 1.04 &10.42 & 1 & 15 & \\     
A1995   & 0.3186 & 4.61 &  4.1 & 0.7 & 24.13 & 0.83 & 8.83 & * & 17 &   \\
A2034   & 0.1130 & 2.03 & 13.6 & 1.0 & 23.64 & 0.61 & 3.81 & * & 16 &   \\
A2163   & 0.2030 & 3.31 &155.0 & 2.0 & 25.26 & 2.28 &22.73 & 7 & 14 & a relic is also present \\
A2218   & 0.1756 & 2.94 &  4.7 & 0.1 & 23.60 & 0.38 & 5.77 & 3 & 16 & asymmetric  \\
A2219   & 0.2256 & 3.58 & 81.0 & 4.0 & 25.08 & 1.72 &12.19 & 1 & 16 &   \\
A2254   & 0.1780 & 2.98 & 33.7 & 1.8 & 24.47 & 0.92 & 4.55 & 2 & 16 & irregular asymmetric shape\\
A2255   & 0.0806 & 1.50 & 56.0 & 3.0 & 23.94 & 0.90 & 2.64 & 8 & 16 & a relic is also present \\
A2256   & 0.0581 & 1.11 &103.4 & 1.1 & 23.91 & 0.81 & 3.75 &9  & 16 & a relic is also present \\
A2294   & 0.1780 & 2.98 &  5.8 & 0.5 & 23.71 & 0.54 & 3.90 & * & 16 &   \\
A2319   & 0.0557 & 1.07 &153.0 & 8.0 & 24.04 & 1.02 & 8.46 &10 & 15 & spectral steepening \\
A2744   & 0.3080 & 4.50 & 57.1 & 2.9 & 25.24 & 1.89 &12.86 & 2 & 21 & a relic is also present \\
A3444   & 0.2533 & 3.91 & 14.6 & 1.0 & 24.45 & 3.3  &13.42 & * & 21 & Filament; see text \\
A3562   & 0.0490 & 0.95 & 20.0 & 2.0 & 23.04 & 0.28 & 1.57 &11 & 15 & spectral steepening \\
1E0657-56&0.2960 & 4.38 & 78.0 & 5.0 & 25.33 & 2.1  &22.59 &12 & 21 & bullet cluster \\
RXCJ1314.4-2515&0.2439&3.81& 20.3&0.8&24.55&1.83  &10.75 &13 & 21 & Double relics are present\\
CL1446+26&0.3700&5.08 &9.2 & 0.5 & 24.63& 0.36    & 3.42 & * & 20 & relic source \\   
RXCJ2003.5-2323&0.3171 &4.59&35.0 &2.0 &25.09 &1.40&9.12  &14 & 21 &  \\
\hline
\multicolumn{11}{l}{\scriptsize Col. 1: Cluster Name; Col. 2: Redshift; Col. 3: 
Angular to linear conversion; Col. 4: Radio flux density at 1.4 GHz;
 Col. 5: Estimated flux density error;}\\
\multicolumn{11}{l}{\scriptsize Col. 6: Radio power at 1.4 GHz; 
Col. 7: Radio largest linear size; 
Col. 8: X-ray luminosity in the 0.1-2.4 keV band in 10$^{44}$ units; }\\  
\multicolumn{11}{l}{\scriptsize Col. 9: References for radio data:}\\
\multicolumn{11}{l}{\scriptsize * = This work; 1 = \cite{bac03}; 
2 = \cite{gov01a}; 3 = \cite{gf00};
 4 = \cite{ven09}; 5 = \cite{rei99};}\\ 
\multicolumn{11}{l}{\scriptsize 6 = \cite{kim90}; 7 = \cite{fer01}; 
8 = \cite{gov05}; 9 = \cite{ce06};}\\ 
\multicolumn{11}{l}{\scriptsize 10 = \cite{fer97};
11 = \cite{ven03}; 12 = \cite{lia00}; 13 = \cite{fer05}; 
14 = \cite{gia09}.}\\
\multicolumn{11}{l}{\scriptsize Col. 10: References for X-ray data:}\\
\multicolumn{11}{l}{\scriptsize 15 = \cite{rb02}; 
16 = \cite{ebe98}; 
17 = \cite{boe00}; 18 = \cite{led03}; 19 = \cite{ebe96}; 
20 = \cite{wu99}; 21 = \cite{boe04}.}\\
\multicolumn{11}{l}{\scriptsize Col. 11: Notes}\\
\end{tabular}
\end{table*}

\subsection{Radio Morphology}

Radio halos are generally extended sources located at the cluster center. 
The largest halo known so far is in A2163 (LLS = 2.28 Mpc), and many halos 
have a projected size larger than 1 Mpc.
We note that the diffuse emission in A3444 is even more extended than A2163, 
but we consider its 3.3 Mpc structure as a giant filament very similar to 
the large scale 
structure seen in the filament of galaxies
ZwCl 2341.1+0000 (\cite{bag02}).
Six radio halos have a size $\ltsim$ 0.5 Mpc. However, the properties of the 
galaxy clusters and the absence of a strong centered radio galaxy indicate
that these sources are very different from the mini-halo
class of radio sources (see e.g. \cite{gov09}).
We consider them as genuine 
halos of small size.

Most radio halos show a centrally located and 
regular radio morphology. In a few cases
the radio source is elongated as e.g. in A209, A401, and A2034; however, we
note that this shape is expected from merging clusters. X-Ray images of merging
clusters can be elongated in the direction of the merger and because
of the strong correlation between the X-ray and radio brightness distribution
(\cite{gov01b}) we expect elongated radio halos. 
Probably the radio emission is still more
complex and irregular with more substructures, as shown e.g. by the 
polarized filament in 
A2255 (\cite{gov05}),
but the large HPBW necessary to detect the
low radio brightness does not allow such a study. 

A few radio halos are highly asymmetric, being located mostly on one side
with respect to the cluster center.
In A851, A1351, A2218, and in the peculiar cluster A1213 we see these 
asymmetric sources.
We interpret these structures as due to a highly asymmetric merger, but a 
more detailed
comparison between radio and X-ray data is necessary.

We note that in 10 clusters with a radio halo 
we confirm the presence of one or two
relic radio 
sources (see notes in Table 2). The high percentage ($\sim$ 30$\%$) of
clusters with both a relic and a halo source confirms the connection between
these two classes of diffuse sources and cluster mergers. In this scenario 
the cluster turbulence arising from 
a cluster merger is the major mechanism
responsible for the electron reacceleration in halo sources, moreover shock
waves originated by major mergers are able to compress magnetic fields and 
reaccelerate electrons at the cluster periphery giving origin to relics.

\subsection{Radio size and Power}

\begin{figure}[!!htb]
\centering
\includegraphics[width=9cm]{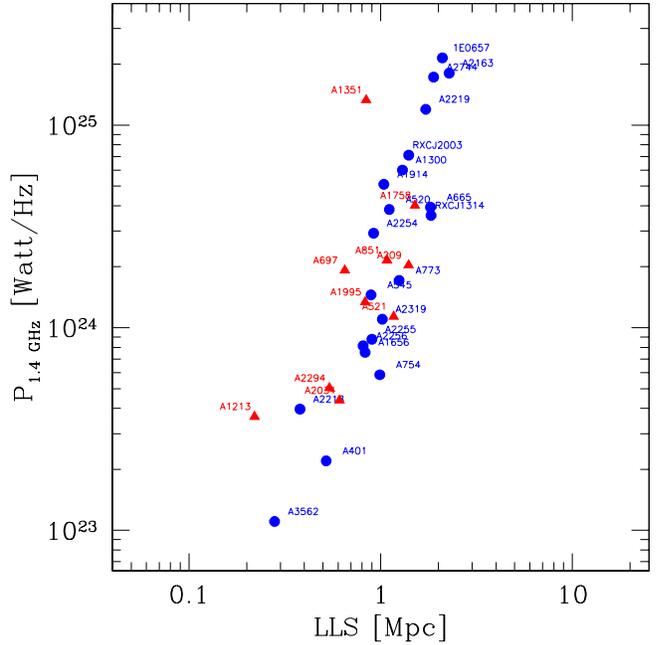}
\caption{Total radio power at 
1.4 GHz versus the largest linear size of radio halos in Mpc, both for new 
halos (red triangles) and halos already known in literature (blue dots).}
\label{radioLLS} 
\end{figure}

In Fig. \ref{radioLLS} we report the radio halo's largest size (LLS) versus 
the total radio power at 1.4 GHz. The LLS has been estimated in radio images at
1.4 GHz at a 3 sigma noise level. 
The data are well correlated in agreement with the
finding by \cite{cas07} and \cite{mur09}. 
Moreover, the relation between the radio halo size and the total
radio power is well-defined also for the smaller radio halos. Sources 
with a total radio power as low as  $\sim$10$^{23}$ W/Hz and a 
LLS $\sim$ 300 kpc
show the same correlation as giant radio halos. This suggest a common origin
and physical mechanism for giant and small size radio halos. 

We note that a more detailed study of this correlation will require a more
detailed measure of the size of radio halos. It is not among the aims of the present
paper to discuss this correlation and we refer to \cite{mur09}
and \cite{cas07} about this topic. Here we want only to point out that
the small dispersion of radio size versus radio power of halos
confirms that giant and smaller radio halos belong 
to the same class of sources and
show similar physical properties. Halos as small as a 
few hundred of kpc show the
same properties of Mpc size giant halos. 

Two sources only are not in good agreement with the radio power - LLS relation:
A1213 and A1351. A1213 as previously discussed is peculiar because of
the low X-Ray Luminosity and the cluster radio properties. A1351 shows
a complex radio morphology therefore deeper observations are needed for this 
cluster.

\subsection{Halo properties and redshift}

\begin{figure}[!!htb]
\centering
\includegraphics[width=9cm]{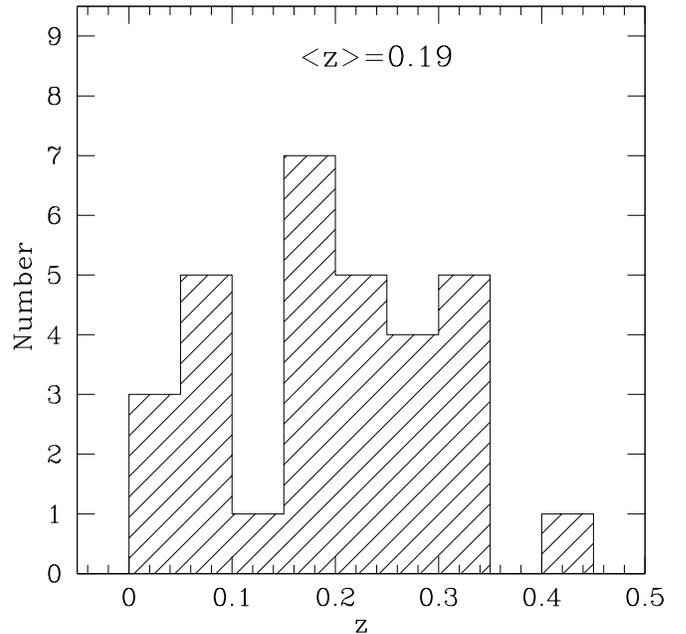}
\caption{Redshift distribution of known radio halos.}
\label{histo} 
\end{figure}

We consider here the redshift distribution of the present sample of radio 
halos.
In Fig. \ref{histo} we show the histogram of the redshift distribution of all 
the halos reported in Tab. 2. 
The redshift distribution is rather homogeneous and there is a relatively 
good statistics to 
investigate halo properties at least up to z = 0.35.
We consider a statistical fluctuation the low number
of halos in between z = 0.1 and 0.15 visible in Fig. \ref{histo}.

\begin{figure}[!!htb]
\centering
\includegraphics[width=9cm]{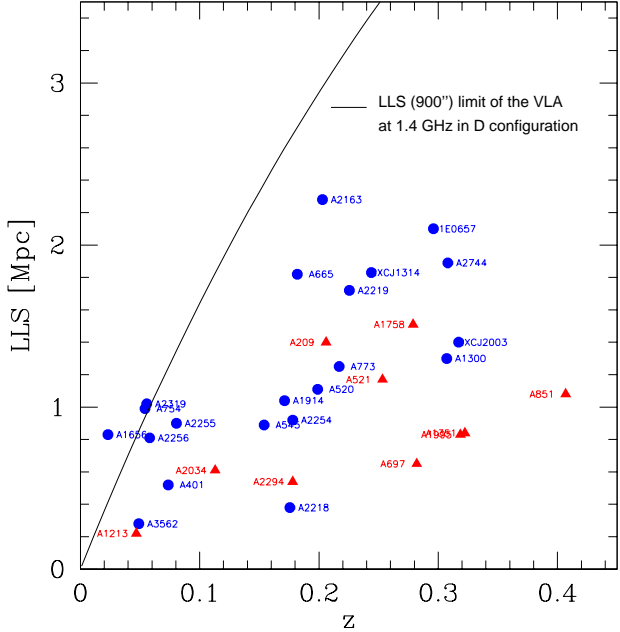}
\caption{Largest Linear size of radio halos in Mpc versus z. The upper line
corresponds to a radio halo with a size = 15', the largest structure
visible by VLA at 1.4 GHz in D configuration. Red triangles are new halos, 
blue dots are halos from literature data.} 
\label{LLSz} 
\end{figure}

To investigate possible selection effects
we show in Fig. \ref{LLSz} the trend of the LLS with the redshift.
We note that most of the halo observations have been obtained with the VLA at
1.4 GHz in the C and D array where the largest visible structure has
an angular size = 15' corresponding to the line drawn in Fig. \ref{LLSz}.
Sources on the left of this line have too large an angular size to be 
recorded by the VLA. 
The few halos with a larger size are: the Coma cluster (A1656), 
not 
visible with the VLA at 1.4 GHz, and studied at this frequency 
by \cite{kim90} combining VLA and DRAO data; A2319 observed by \cite{fer97}
with the WSRT; A754 visible in VLA images because of the
elongated and irregular structure of the radio halo.

From this distribution we can derive an upper limit of about 2 Mpc for the halo
size, probably related to the physical dimension of galaxy clusters, and
the gas distribution. The lack of large size halos at high redshift is 
apparent since we have
only 1 halo at z $>$ 0.35. 

\begin{figure}[!!htb]
\centering
\includegraphics[width=9cm]{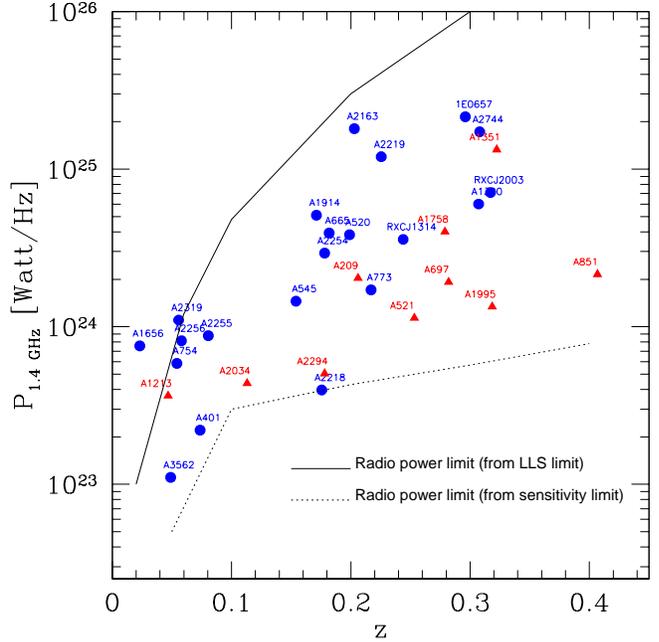}
\caption{Total radio power at 1.4 GHz versus z. The upper continuous line
corresponds to a halo linear size corresponding to 15', the upper limit
for VLA observations at 1.4 GHz. The lower dotted line is from an average 
sensitivity limit assuming a standard VLA observation with an integration 
time of $\sim$ 3 hrs (see text). Red triangles are new halos, 
blue dots are halos from literature data.}
\label{radioz} 
\end{figure}

\begin{table*}
\caption{Spectral Index of Radio halos}
\label{tab3}
\centering
\begin{tabular}{lcccc}
\hline\hline
Name     & $\alpha$1     & $\alpha$2 & Temperature & Reference \\
         &               &           &  keV        &           \\
\hline
\hline
Straight & $>$ 3 points                 &    &  &           \\
A1914    & $\alpha_{0.26}^{1.4}$ = 1.88 &    &  7.9 & 1        \\
A2256    & $\alpha_{0.22}^{1.4}$ = 1.61 &    &  6.6 & 7       \\
1E 0657  & $\alpha_{0.84}^{5.9}$ = 1.3  &    &  10.6 & 10       \\
\hline
Steepening& $>$ 3 points                &           \\
A1656    & $\alpha_{0.31}^{1.4}$ = 1.16 &$\alpha_{1.4}^{4.8}$ = 2.28& 8.4&11 \\
\hline
\hline
Straight & 3 points                &   &         &  \\
A521     & $\alpha_{0.24}^{1.4}$ = 1.80 &    &  5.9 & *           \\
RXC J2003& $\alpha_{0.2}^{1.4}$ = 1.3 &    & - & 4       \\
\hline
Steepening& 3 points                &         &   \\
A754     & $\alpha_{0.07}^{0.3}$ = 1.1 &$\alpha_{0.3}^{1.4}$ = 1.5 & 9.5 & 1 \\
A2163    & $\alpha_{0.3}^{1.3}$  = 1.1 &$\alpha_{1.3}^{1.6}$ = 1.5 &13.3 & 5 \\
A2319    & $\alpha_{0.4}^{0.6}$  = 0.9 &$\alpha_{0.6}^{1.4}$ = 2.2 & 8.8 &8 \\
A3562    & $\alpha_{0.3}^{0.8}$  = 1.3 &$\alpha_{0.8}^{1.4}$ = 2.1 & 5.2 &9 \\
\hline
\hline
          & 2 points               &   & \\
A545     & $\alpha_{1.3}^{1.6}$ $>$ 1.4 &   & 5.5 & 1       \\
A665     & $\alpha_{0.3}^{1.4}$ = 1.04 &    & 9.0 & 3       \\
A697     & $\alpha_{0.3}^{1.4}$ = 1.2  &    & 10.2& *       \\
A1300    & $\alpha_{0.3}^{1.4}$ = 1.8 &     & 9.2 & 4      \\
A2218    & $\alpha_{1.4}^{5.0}$ = 1.6 &     & 7.1 & 3      \\
A2219    & $\alpha_{0.3}^{1.4}$ = 0.9  &    & 12.4 & 2       \\
A2255    & $\alpha_{0.3}^{1.4}$ = 1.7 &     & 6.9 & 6      \\
A2744    & $\alpha_{0.3}^{1.4}$ = 1.0 &     & 10.1 & 2      \\
\hline
\multicolumn{4}{l}{\scriptsize Notes: Temperature is the average cluster 
Temperature in keV from literature data;}\\
\multicolumn{4}{l}{\scriptsize References to spectral index information:}\\
\multicolumn{4}{l}{\scriptsize
* = This work; 1 = \cite{bac03}; 
2 = \cite{orr07}; }\\
\multicolumn{4}{l}{\scriptsize
3 = \cite{gf00};
4 = \cite{ven09};}\\
\multicolumn{4}{l}{\scriptsize  
5 = \cite{fer01}; 6 = \cite{gov05};}\\
\multicolumn{4}{l}{\scriptsize 
7 = \cite{bre08}; 
8 = \cite{fer97};}\\
\multicolumn{4}{l}{\scriptsize 
9 = \cite{ven03};
10 = \cite{lia00}; 11 = \cite{thi03}.}\\
\end{tabular}
\end{table*}

In Fig. \ref{radioz} we show the trend of radio power with redshift.
To discuss possible selection effects in the halo distribution shown 
in this figure we individuate
the regions where it is possible to obtain radio data with
present radio telescopes.
The upper continuous line has been derived taking into account the largest 
angular size
visible with the VLA at 1.4 GHz (as discussed in Fig. \ref{LLSz}),
and the correlation between size and radio
power shown in Fig. \ref{radioLLS}. Halo sources in the region above the
continuous line are not visible because their angular size is too large.
The lower dotted line has been estimated assuming that most of the data are
from relatively short VLA observations. We have considered a $\sim$ 3 hours
observing time. Therefore radio halos in the region below the dotted 
line are not present because of sensitivity limits. 
To derive this limit we have taken into account the brightness
decrease because of the luminosity and 
angular size distance and the correlation between 
radio power and LLS.
Clear selection effects are present. With available radio telescopes
we can observe only radio halos in between the two lines, BUT we note that 
radio halos discussed here are distributed homogeneously in the allowed region.
The present sample is not complete, but it is 
representative
of radio halos that can be observed with present radio telescopes.

More sensitive observations with the VLA and the EVLA should allow us to detect
possible fainter sources. We will need to wait for 
the new generation of radio
telescopes (LOFAR, SKA-pathfinders, SKA) to search for diffuse sources
outside the present allowed region. 

\subsection{Spectral index}

The integrated radio spectrum of halo sources is poorly known. These
extended and diffuse sources always show a steep spectrum but only in a few 
cases the spectrum is derived with more than three flux density 
measurements at 
different frequencies. Moreover for most sources the highest 
available frequency
is 1.4 GHz and the presence of a spectral steepening, crucial to discriminate 
between different reacceleration models, is difficult to determine.

Only four radio halos have a really well defined radio spectrum. 
The best studied integrated spectrum is still that for the Coma
cluster where a clear evidence of a high frequency steepening is present
(see \cite{thi03} for a detailed discussion).
The steepest straight spectrum is shown by the radio halo in A1914 (\cite{bac03}) 
where 9 different points show a straight spectrum with
$\alpha$ = 1.88. One more well studied halo is the one in A2256 
(\cite{bre08}), and in
1E 0657 discussed in detail by \cite{lia00}.

\cite{fer04b} investigated the existence of a possible correlation
between the spectral index of radio halos and the cluster temperature.
They found a marginal evidence that clusters at higher temperature tend to
host halos with a flatter spectrum. 
To investigate this correlation we divided 
the radio halos with spectral information (Table 3) into three groups
to enhance the statistical differences taking into account that
the spectral index measures are highly inhomogeneous. 
For radio halos with a steepening spectrum we considered the average value
in the range between 0.3 and 1.4 GHz. Because of the short frequency range we 
did not include A545.
We obtained the following 
result:

\begin{itemize}
\item
radio halos in clusters with an average temperature less than 8 keV, show
an average spectral index of 1.7; 
\item
radio halos in clusters 
with a temperature in the range from 8 to 10 keV, show an average spectral 
index of 1.4; 
\item
halos in clusters with a temperature greater than 10 keV show an average 
spectral index of 1.1.
\end{itemize}
Despite the uncertainties due to highly inhomogeneous spectral index 
measures, 
this result indicates that hotter cluster tend to host halos with flatter 
spectra. This may be understood in the framework of acceleration models, since
the hottest clusters are more massive and may host more recent violent mergers
able to reaccelerate 
relativistic particles giving origin to halos with a flatter radio spectrum.

Note that in Table 3 we do not report uncertainties on the spectral index
value, because: i) in well known radio spectra (more than 3 points) the
spectral index is well determined, with an uncertainty due to the 
best fit procedure (usually of the order of 0.01); ii) in other radio
spectra with only 2 or 3 points the formal error on the spectral index
value is formally of the order of 0.1, {\it but} the largest uncertainty
is the possible presence of a change in the radio spectrum curvature (e.g. steepening)
which cannot be seen with only 2 - 3 points. We think that this uncertainty is
minimized by our statistical 
approach to separate radio halos with spectral index information
in 3 groups.

\subsection{Radio X-ray correlation}

\begin{figure}[!!htb]
\centering
\includegraphics[width=9cm]{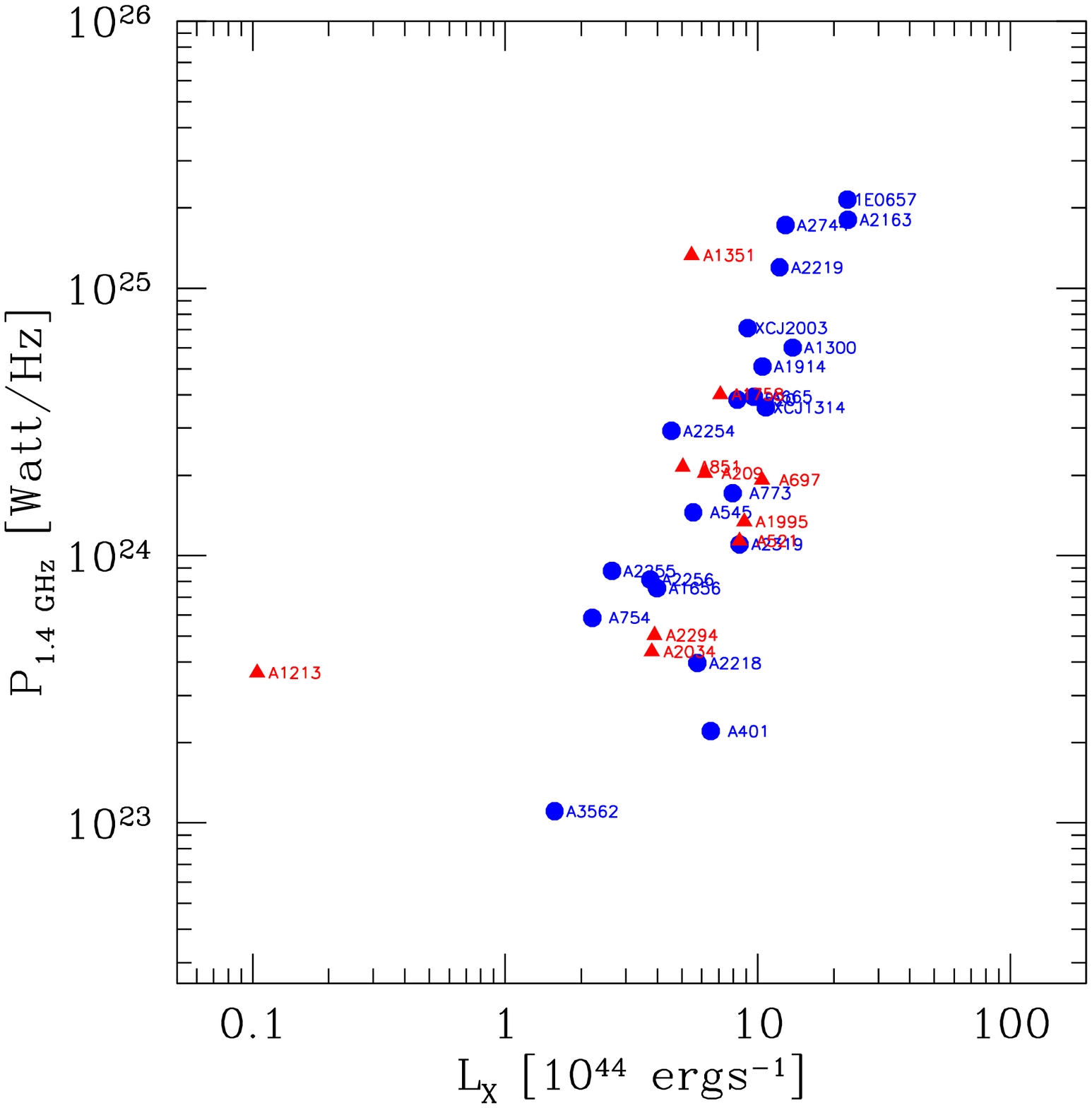}
\caption{Cluster X-Ray Luminosity between 0.1 and 2.4 keV versus the total 
halo radio power at 1.4 GHz. Red triangles are new halos, 
blue dots are halos from literature data.}
\label{radioX} 
\end{figure}

In Fig. \ref{radioX} we present the relation between the halo radio power 
at 1.4 GHz and the X-ray Luminosity. The distribution of present data is
in agreement with previous results of this well known relation (see e.g. 
\cite{bac03}, \cite{fer05b}), but
we have now a larger sample with a larger range in X-ray Luminosity and 
radio power.
We note that this correlation is not generally applicable to all clusters
but 
is present only between radio halos and parent galaxy clusters:
relaxed, cooling clusters have not to be considered for this correlation. 

The only cluster outside the correlation is Abell 1213 whose peculiarity has
been already discussed in Sects. 3 and 4.2. It is the cluster in our
sample with the lowest X-ray Luminosity (0.10 $\times$ 10$^{44}$ erg/sec
in the 0.1 - 2.4 keV band (see Table 2). The diffuse radio emission is 
the smallest (220 kpc) source discussed here, but because of its low redshift
it is well resolved in NVSS images and confirmed by pointed observations 
obtained by us.
This cluster could belong to a 
poorly known class of objects where a diffuse radio emission is present
despite the very low X-ray emission.

\section{Conclusion}

We have presented data on all the radio halos
in clusters at z $<$ 0.4 observed up to now.
Adding new results (from proprietary and archive VLA data) to 
published data,
we collected an homogeneous sample of 31 radio halos. 
Data at 1.4 GHz are available
for all sources. From a study of this large sample we have concluded that:

\begin{itemize}

\item
the largest halo known is in A2163 with a LLS = 2.28 Mpc. Larger halos
should be visible with present radio telescopes, but have not been found.
We suggest that an upper limit on the radio halo size is expected because of
the limited size of galaxy clusters and merger events. 
A few halos have a size $\ltsim$ 0.5 Mpc,
but their properties are similar to those of giant halos

\item
most radio halos are centrally located and show a regular shape. 
In a few cases the
radio structure is irregular and elongated. In particular in 4 clusters the 
radio emission is centrally located, but extended only 
on one side with respect to the
cluster center. We interpret these 
structures as related to smaller and asymmetric cluster mergers

\item
in $\sim$ 30$\%$ of clusters with a radio halo a relic
radio source is  also present, in agreement with the common relation between 
these two classes 
of sources and merger events

\item
we find a good correlation between radio size and power for small
and giant radio halos. This result suggests a common origin and similar 
physical properties in small size and giant radio halos.

\item
from the power and size distribution with redshift of radio halos we derive 
that known radio halos are homogeneously distributed in the observable
region covered by the present radio telescopes. To investigate the
presence of possible {\it different} radio halos, new instruments such as
the Low Frequency ARray (LOFAR - NL), the Low Wawelength Array (LWA - US),
the Australian Square Kilometre Array Pathfinder (ASKAP), 
and the Square Kilometer Array (SKA)
are necessary. However we point out that observations with a better sensitivity
as possible in the near future with the EVLA, could improve our knowledge
of small and faint radio halos

\item
we confirm with a better statistics that the radio spectra of halos
are related to the cluster temperature, being flatter in hotter cluster.
Since the cluster temperature is a good indication of the
turbulence
present in the ICM, this correlation favours the interpretation
that turbulence is the mechanism responsible to
supply energy to relativistic electrons

\item
we confirm with better statistics the correlation between cluster X-ray 
luminosity and radio power. We note the exception of the peculiar cluster
A1213 where a diffuse radio emission is clearly present but the X-ray
Luminosity is very low. We suggest that A1213 and a few other cases discussed 
in the literature (see e.g. \cite{br09}) could represent a class of 
objects with an extended radio 
emission not clearly related to the cluster X-ray properties. Because of the
small number of these sources we cannot yet discuss their properties, and
we are aware that their existence is to be confirmed.
However we note that any
alternative explanations for the diffuse emission in Abell 1213 (extended 
head-tail morphology of source B and/or 4C29.41) are less convincing.

\item
among new diffuse sources presented here there is also a new relic 
(in CL1446+26) and 
a giant filamentary emission extended more than 3 Mpc 
(in A3444) which will be discussed
in detail in the future.

\end{itemize}

\begin{acknowledgements}
We thank J. Morgan for a critical reading of the manuscript and the
anonymous Referee for useful comments and suggestions.
The National Radio Astronomy
Observatory is operated by Associated Universities, Inc., under cooperative
agreement with the National Science Foundation.
This research has made use of
the NASA/IPAC Extragalactic Data Base (NED) which is operated by the JPL,
California Institute of Technology, under contract with the National
Aeronautics and Space Administration.
This research was partially
supported by ASI-INAF I/088/06/0 - High Energy Astrophysics.
\end{acknowledgements}


\begin{thebibliography}{}

\bibitem[Bacchi et al. 2003]{bac03} 
Bacchi M., Feretti L., Giovannini G., Govoni, F.\ 2003, A\&A, 400, 465 

\bibitem[Bagchi et al. 2002]{bag02} 
Bagchi, J., En{\ss}lin, T.~A., Miniati, F., et al.,\ 2002, New Astronomy, 7, 249 

\bibitem[Becker et al. 1995]{bec95}
Becker, R.~H., White, R.~L., \& Helfand, D.~J. 1995, \apj, 450, 559

\bibitem[B{\"o}hringer et al. 2000]{boe00} 
B{\"o}hringer, H., Voges, W., Huchra, J.P., et al.\ 2000, \apjs, 129, 435 

\bibitem[B{\"o}hringer et al. 2004]{boe04} 
B{\"o}hringer, H., Schuecker, P., Guzzo, L., et al.\ 2004, \aap, 425, 367 

\bibitem[Bonafede et al. 2009a]{bon09a} 
Bonafede, A., Giovannini, G., Feretti, et al.\ 2009a, \aap, 494, 429 

\bibitem[Bonafede et al. 2009b]{bon09b} 
Bonafede, A., Feretti, L., Giovannini, G., Govoni, F., Murgia, M.,
Taylor, G.B., Ebeling, H., Allen, S., Gentile, G., Pihlstrom, Y.\ 2009b,
\aap, in press - arXiv:0905.3552

\bibitem[Brentjens 2008]{bre08} 
Brentjens, M.A. \ 2008, \aap, 489, 69

\bibitem[Brown \& Rudnick 2009]{br09} 
Brown, S., \& Rudnick, L. \ 2009, \aj, 137, 3158

\bibitem[Brunetti et al. 2001]{bru01} 
Brunetti, G., Setti, G., Feretti, L., \& Giovannini, G.\ 2001, \mnras, 320, 365 

\bibitem[Brunetti et al. 2008]{bru08} 
Brunetti, G., Giacintucci, S., Cassano, R., et al.\ 2008, \nat, 455, 944 

\bibitem[Buote 2001]{buo01} 
Buote, D.~A.\ 2001, \apjl, 553, L15 

\bibitem[Cassano et al. 2007]{cas07} 
Cassano, R., Brunetti, G., Setti, et al.\ 2007, \mnras, 378, 1565 

\bibitem[Clarke \& Ensllin 2006]{ce06} 
Clarke, T.~E., \& Ensslin, T.~A.\ 2006, \aj, 131, 2900 

\bibitem[Dahle et al. 2002]{dah02} 
Dahle, H., Kaiser, N., Irgens, R.~J., et al.\ 2002, \apjs, 139, 313 

\bibitem[Dallacasa et al. 2009]{dal09} 
Dallacasa, D., Brunetti, G., Giacintucci, S., Cassano, R., Venturi, T.,
Macario, G., Kassim, N.E., Lane, W., Setti, G.\ 2009, \apj, in press - 
arXiv:0905.0588 

\bibitem[David \& Kempner 2004]{dk04} 
David, L.~P., \& Kempner, J.\ 2004, \apj, 613, 831 

\bibitem[Ebeling et al. 1996]{ebe96} 
Ebeling, H., Voges, W.,  B{\"o}hringer, H. et al.\ 1996, \mnras, 281, 799 

\bibitem[Ebeling et al. 1998]{ebe98} 
Ebeling, H., Edge, A.~C., B{\"o}hringer, H., et al.\ 1998, \mnras, 301, 881 

\bibitem[Fanti et al. 1982]{fan82} 
Fanti, C., Fanti, R., Feretti, L., et al.\ 1982, \aap, 105, 200 

\bibitem[Feretti et al. 1997]{fer97} 
Feretti, L., Giovannini, G., \&  B{\"o}hringer, H.\ 1997, New Astronomy, 2, 501 
\bibitem[Feretti \& Giovannini 1998]{fg98} 
Feretti, L., \& Giovannini, G.\ 1998, Untangling Coma Berenices: 
A New Vision of an Old Cluster, 123 

\bibitem[Feretti 2000]{fer00} 
Feretti, L.\ 2000, Invited review at IAU 199 `The Universe at Low Radio Frequencies' in Pune, India, 1999,
arXiv:astro-ph/0006379 

\bibitem[Feretti et al. 2001]{fer01} 
Feretti, L., Fusco-Femiano, R., Giovannini, G., \& Govoni, F.\ 2001, \aap, 373, 106 

\bibitem[Feretti 2004]{fer04a} 
Feretti, L.\ 2004, 35th COSPAR Scientific Assembly, p.1425

\bibitem[Feretti et al. 2004]{fer04b} 
Feretti, L., Brunetti, G., Giovannini, G., Kassim, N., Orru', E., Setti, G.
\ 2004b Journ. of the Korean Astron. Soc. 37, 315

\bibitem[Feretti 2005a]{fer05a} 
Feretti, L.\ 2005a, Advances in Space Research, 36, 729 

\bibitem[Feretti 2005b]{fer05b} 
Feretti, L.\ 2005b, X-Ray and Radio Connections (eds. L.O. Sjouwerman and K.K 
Dyer) Published electronically by NRAO, http://www.aoc.nrao.edu/events/xraydio
(arXiv:astro-ph/0406090)

\bibitem[Feretti et al. 2005]{fer05} 
Feretti, L., Schuecker, P., B{\"o}hringer, H., et al.,\ 2005 \aap, 444, 157 

\bibitem[Feretti \& Giovannini 2008]{fg08} 
Feretti, L., \& Giovannini, G.\ 2008, A Pan-Chromatic View of Clusters of
Galaxies and the Large-Scale structure, edited by M. Plionis, O. López-Cruz 
and D. Hughes. Lecture Notes in Physics Vol. 740.474, p.143 

\bibitem[Ferrari et al. 2006]{ferra06} 
Ferrari, C., Arnaud, M., Ettori, S., et al.,\ 2006, \aap, 446, 417 

\bibitem[Ferrari et al. 2008]{ferra08} 
Ferrari, C., Govoni, F., Schindler, S., Bykov, 
A.M., \& Rephaeli, Y.\ 2008, Space Science Reviews, 134, 93 

\bibitem[Giacintucci et al. 2006]{gia06} 
Giacintucci, S., Venturi, T., Bardelli, S., et al.,\ 2006, New Astronomy, 11, 437 

\bibitem[Giacintucci et al. 2008]{gia08} 
Giacintucci, S., Venturi, T., Macario, G., et al.\ 2008, \aap, 486, 347 

\bibitem[Giacintucci et al. 2009]{gia09} 
Giacintucci, S., Venturi, T., G. Brunetti, et al.\ 2009, \aap, in press;
arXiv:0905.3479

\bibitem[Giovannini et al. 1999]{gio99}
Giovannini, G., Tordi, M., Feretti, L. \ 1999, NewA, 4, 141

\bibitem[Giovannini \& Feretti 2000]{gf00} 
Giovannini, G., \& Feretti, L.\ 2000, New Astronomy, 5, 335 

\bibitem[Giovannini \& Feretti 2004]{gf04} 
Giovannini, G., \& Feretti, L.\ 2004, Journal of Korean Astronomical Society, 37, 323 

\bibitem[Giovannini et al. 2006]{gio06} 
Giovannini, G., Feretti, L., Govoni, F., et al.,\ 2006, Astronomische Nachrichten, 327, 563 

\bibitem[Girardi et al. 2006]{gir06} 
Girardi, M., Boschin, W., \& Barrena, R.\ 2006, \aap, 455, 45 

\bibitem[Govoni et al. 2001a]{gov01a} 
Govoni, F., Feretti, L., Giovannini, et al.,\ 2001a, \aap, 376, 803 

\bibitem[Govoni et al. 2001b]{gov01b} 
Govoni, F., Ensslin, T.A., Feretti, L., Giovannini, et al.,\ 2001b, \aap, 
369, 441

\bibitem[Govoni \& Feretti 2004]{govf04} 
Govoni, F., \& Feretti, L.\ 2004, International Journal of Modern Physics D, 13, 1549 

\bibitem[Govoni et al. 2004]{gov04} 
Govoni, F., Markevitch, M., Vikhlinin, A., et al.,\ 2004, \apj, 605, 695 

\bibitem[Govoni et al. 2005]{gov05} 
Govoni, F., Murgia, M., Feretti, L., et al.,\ 2005, \aap, 430, L5 

\bibitem[Govoni et al. 2009]{gov09} 
Govoni, F., Murgia, M., Markevitch, M. et al.,\ 2009, A\&A, 499, 371

\bibitem[Jones \& Forman 1999]{jf99} 
Jones, C., \& Forman, W.\ 1999, \apj, 511, 65 

\bibitem[Kassim et al. 2001]{kas01} 
Kassim, N.~E., Clarke, T.~E., En{\ss}lin, T.~A., et al., 2001, \apj, 559, 785 

\bibitem[Kempner \& Sarazin 2001]{ks01} 
Kempner, J.~C., \& Sarazin, C.~L.\ 2001, \apj, 548, 639 

\bibitem[Kempner et al. 2003]{kem03} 
Kempner, J.~C., Sarazin, C.~L., \& Markevitch, M.\ 2003, \apj, 593, 291 

\bibitem[Kim et al. 1990]{kim90}
Kim, K-T., Kronberg, P. P., Dewdney, P. E., Landecker, T. L.\ 1990, \apj,
355, 29

\bibitem[Liang et al. 2000]{lia00} 
Liang, H., Hunstead, R.~W., Birkinshaw, M., \& Andreani, P.\ 2000, \apj, 544, 686 

\bibitem[Ledlow et al. 2003]{led03} 
Ledlow, M.~J., Voges, W., Owen, F.~N., \& Burns, J.~O.\ 2003, \aj, 126, 2740 

\bibitem[Lemonon (1999)]{lem99}
Lemonon, L.\ 1999, Ph.D. Thesis Universite' de Paris XI

\bibitem[Matsumoto et al. 2001]{mat01} 
Matsumoto, H., Pierre, M., Tsuru, T.~G., \& Davis, D.~S.\ 2001, \aap, 374, 28 

\bibitem[Mercurio et al. 2003]{mer03} 
Mercurio, A., Girardi, M., Boschin, W., et al.,\ 2003, \aap, 397, 431 
 
\bibitem[Murgia et al. 2009]{mur09} 
Murgia, M., Govoni, F., Markevitch, M., et al.,\ 2009, A\&A, 499, 679 

\bibitem[Orru' et al. 2007]{orr07} 
Orru', E., Murgia, M., Feretti, L., Govoni, F., Brunetti, G., Giovannini, G.,
et al.,\ 2007, \aap, 467, 943 

\bibitem[Owen et al. 1999]{owe99} 
Owen, F., Morrison, G., \& Voges, W.\ 1999, Diffuse Thermal and Relativistic Plasma in Galaxy Clusters, 9 

\bibitem[Patel et al. 2000]{pat00} 
Patel, S.~K., Marshall, J., Carlstrom, J.E., et al.\ 2000, \apj, 541, 37

\bibitem[Paulin-Henriksson et al. 2007]{p-h07} 
Paulin-Henriksson, S., Antonuccio-Delogu, V., Haines, C.P., et al., \ 2007,
 \aap, 467, 427

\bibitem[Petrosian 2001]{pet01} 
Petrosian, V.\ 2001, \apj, 557, 560 

\bibitem[Reid et al. 1999]{rei99} 
Reid, A.~D., Hunstead, R.~W., Lemonon, L., \& Pierre, M.~M.\ 1999, \mnras, 302, 571 

\bibitem[Reiprich \& B{\"o}hringer 2002]{rb02} 
Reiprich, T.~H., B{\"o}hringer, H.\ 2002, \apj, 567, 716 

\bibitem[Rizza et al. 1998]{riz98} 
Rizza, E., Burns, J.~O., Ledlow, et al.,\ 1998, \mnras, 301, 328 

\bibitem[Sato \& Martin 2006]{sm06} 
Sato, T., \& Martin, C.~L.\ 2006, \apj, 647, 946 

\bibitem[Thierbach et al. 2003]{thi03} 
Thierbach, M., Klein, U., \& Wielebinski, R.\ 2003, \aap, 397, 53 

\bibitem[Venturi et al. 2003]{ven03} 
Venturi, T., Bardelli, S., Dallacasa, D., et al..\ 2003, \aap, 402, 913 

\bibitem[Venturi et al. 2007]{ven07} 
Venturi, T., Giacintucci, S., Brunetti, G., et al.\ 2007, \aap, 463, 937 

\bibitem[Venturi et al. 2008]{ven08} 
Venturi, T., Giacintucci, S., Dallacasa, D., et al.\ 2008, \aap, 484, 327 

\bibitem[Venturi et al. 2009]{ven09} 
Venturi, T., Giacintucci, S., Cassano, R., et al.\ 2009, "The Low Frequency Radio Universe", ASP Conference Series, arXiv:0903.2934 

\bibitem[Wu et al. 1999]{wu99}
Wu, X.-P., Xue, Y.-J., \& Fang, L.-Z.\ 1999, \apj, 524, 22

\end{thebibliography}
\end{document}